\newcommand{\be}{\begin{equation}}
\newcommand{\ee}{\end{equation}}
\newcommand{\bea}{\begin{eqnarray}}
\newcommand{\eea}{\end{eqnarray}}
\newcommand{\ba}[1]{\begin{array}{#1}}
\newcommand{\ea}{\end{array}}
\documentclass[onecolumn,secnumarabic,amssymb,nobibnotes,nofootinbib,aps,pra,showpacs]{revtex4}
\usepackage{epsfig}
\usepackage{amsmath}
\usepackage{amssymb}
\usepackage{mathrsfs}
\usepackage{color}
\usepackage{rotating}

\begin{document}
\setlength{\topmargin}{-0.5in}

\title{Fermionic vs. bosonic two-site Hubbard models with a pair of interacting cold atoms}
\author{Subhanka Mal$^1$, Kingshuk Adhikary$^1$, and Bimalendu Deb$^{1,2}$}
\affiliation{$^1$Department of Materials Science, 
$^2$Raman Center for Atomic, Molecular and Optical Sciences,   Indian Association
for the Cultivation of Science,
Jadavpur, Kolkata 700032, India.}
\def\zbf#1{{\bf {#1}}}
\def\bfm#1{\mbox{\boldmath $#1$}}
\def\hf{\frac{1}{2}}
\begin{abstract}
In a recent work, Murmann {\it et. al.} [Phys. Rev. Lett. {\bf114}, 080402 (2015)] have experimentally prepared and manipulated a double-well optical potential containing a pair of Fermi atoms as a possible building block of Hubbard model. Here, we carry out a detailed theoretical study on the properties of both fermionic and bosonic two-site Hubbard models with a pair of interacting atoms in a trap with a double-well structure along z-axis and a 2D harmonic confinement along the transverse directions. We consider fermions as of two-component type and bosons as of spinless as well as of two spin components. We first discuss building up the Hubbard models using the model finite-range interaction potentials of Jost and Kohn. In general, a finite range of interaction leads to on-site, inter-site, exchange and partial-exchange terms. We show that, given the same input parameters for both bosonic and fermionic two-site Hubbard models, many of the statistical properties such as the single- and double-occupancy of 
a site, and the probabilities for the single-particle and pair tunneling  are similar in both fermionic and bosonic cases. But, quantum entanglement and quantum fluctuations are found to be markedly different for the two cases. We discuss atom-atom entanglement in two spatial modes corresponding to the two sites of the double-well. Our results show that the entanglement of a pair of spin-half fermions is always greater than that of spinless bosons; and when the fermions are maximally entangled the fluctuation in the two-mode phase difference is largely squeezed. In contrast, spinless bosons never exhibit phase squeezing, but shows squeezing in two-mode population imbalance depending on the system parameters.
\end{abstract}

\maketitle
\section{Introduction}\label{1}

Ultracold atoms in optical lattices have become a testing ground for quantum many-body physics. In this context, a paradigmatic model is the Hubbard model \cite{hubbard:original}, introduced more than fifty years ago to describe the behavior of strongly correlated electrons, and  Mott-insulator transition \cite{Hubbard:1964} in crystalline solids. After the discovery of high-temperature superconductivity in cuprate solids in 80s, it is believed that the model can capture some of the essential aspects of such superconducting phase. In late 80s and early 90s, a bosonic version of the model was formulated  \cite{bose-hubbard} to account for superfluid-to-Mott insulator transition in bosonic lattice systems. With the recent advent of laser-generated optical lattices that provide a pristine crystalline structure for ultracold atoms, both Fermi- and Bose-Hubbard models have attracted renewed interests \cite{Bloch:RMP}, enabling experimenters to realize atomic Bose-Hubbard model \cite{Jaksch:1998}, to demonstrate 
superfluid-Mott insulator transition of Bose-condensed atoms \cite{Greiner:2002}, Fermi surfaces and Fermi-Hubbard model for ultracold fermionic atoms \cite{Esslinger:2008,Bloch:2008}, etc.

One can engineer optical lattice structure with a lot of control over its parameters by external fields. Moreover, the interactions between atoms can be tuned by magnetically controlled Feshbach resonances\cite{chin:rmp:2010}, unlike those between electrons in solids. These features make an optical lattice a possible quantum simulator for many-body quantum systems - a long-sought goal first theoretically envisioned  by Feynman \cite{feynman}. Towards this endeavor, a unique system is the optically or magneto-optically generated double wells or double-well (DW) lattices \cite{Strabley} which have enabled  experimental realizations of a number of correlation effects such as  highly controllable second order tunneling \cite{2ndtunn} and entanglement between isolated atom pairs \cite{nature:phillips,Foot}. About a decade ago, Bloch's group experimentally realized a two-site version of Bose-Hubbard model with a pair of bosonic atoms in two different spin states in a DW and thus demonstrated time-resolved 
 controlled superexchange interaction \cite{supexch}. Recently,  Murmann {\it et al.} \cite{Murmann} have shown a crucial step towards realizing Hubbard model from a bottom-up approach, by preparing  and controlling the quantum states of a pair of interacting two-component fermionic $^6$Li atoms  in a single double-well optical micro-potential. A DW trap loaded with ultracold atoms under tight-binding approximation is considered as a two-site Hubbard model \cite{2ndtunn,Murmann} - a possible building block for creating a full-fledged Hubbard model form a bottom-up approach. Over the last few years, a pair of two-component fermions or bosons in a one-dimensional (1D) DW trap has been employed by several groups \cite{Nano:landman,PRA:carvalho:2018,Dobrzyniecki:2016} for exploring numerous aspects of the two-site model such as spatial and momentum correlations  with ``wave-function anatomy'' of the atom-pair \cite{landman:pra:2018}, fermionization limit of strongly interacting bosonic system \cite{Schmelcher:PRL:2008}, an atomic analog of Hong-Ou-Mandel effect \cite{hong_ou_mandel_PRL:2018} and so on. However, it is not a priori clear how the results obtained and insight gained from the studies with a pair of two-component fermions will change if the fermionic pair is replaced by a pair of spin-polarized or spinless or two-component bosons.

The purpose of this paper is two-fold. First, to understand the efficacy of building up a two-site Hubbard model with a pair of interacting atoms confined in a 3D trap having a DW structure along the one direction. Second, to carry out a detailed model study on the physical and dynamical properties of the system and thereby to compare bosonic and fermionic two-site Hubbard models. We consider fermions as having two spin components and the bosons as spinless or spin-polarized as well as of two spin components. We assume that, unlike commonly used contact-type pseudo-potential, the atoms interact via finite-range model interaction potentials of Jost and Kohn \cite{JostKhon1,JostKhon,our_arxiv_pap}. We use exact numerical single-particle solutions of the trap in calculating the Hubbard parameters under tight-binding two-mode approximation. We also calculate the full six dimensional wave function of the trapped atom-pair including the  interaction potential and use this wave function to calculate the Hubbard interaction parameters which are then compared with the former ones.    
 We find that finite-range of interaction gives rise to three additional interaction parameters apart from the usual on-site interaction $U$. These parameters are the inter-site interaction  $U_i$, the exchange interaction term $K$ and the partial exchange term $I$. We also calculate the on-site interaction term using two-particle wave-functions of two interacting particles. 
 \begin{figure}[h]
    \centering
        \includegraphics[width=4.8in, height=3.2in]{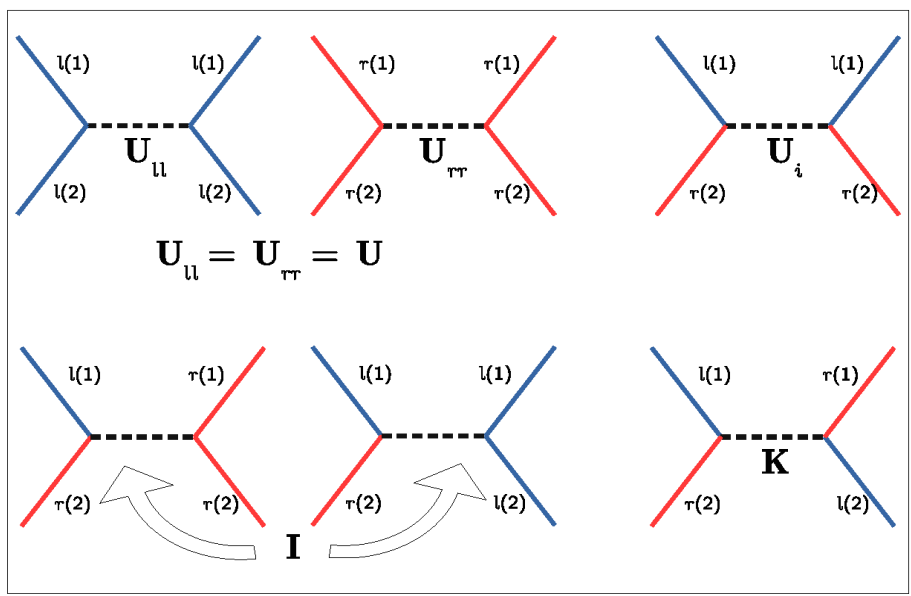}
 \caption{\small Schematic diagrams showing four types of interactions. Here $l(j)$ ($r(j))$ represents $j$th atom in left (right) well. }
 \label{Figure 1.}
\end{figure}
 Our analytical and numerical results show that, for the same input parameters, the quantum-statistically averaged quantities such as single and double occupancy of a site, and the probabilities for the single-particle and pair tunneling are identical or qualitatively similar for a pair of  
 spinless bosons  and a pair of two-component fermions. However, the properties of quantum entanglement and quantum fluctuations are found to be markedly different for a pair of spin-half fermions vis-a-vis a pair of spinless bosons. In terms of number and quantum phase variables, the number and phase squeezing properties are also different in two cases \cite{jpb:2018:kingshuk}.  Quantum phase fluctuations are calculated using the recently introduced quantum mechanical phase operators for matter-waves \cite{jpb:2013:biswajit}.
  
The paper is organized in the following way. 
In the next section,  we discuss how to build up the basic ingredients of a two-site Hubbard model, namely the on-site and other interaction matrix elements, starting from two interacting cold atoms in a DW potential. In Sec.\ref{3}, we discuss two-site bosonic and fermionic Hubbard models. We present and discuss numerical results concerning static and dynamic properties of the two models in Sec.\ref{4}. We describe the properties of quantum entanglement and quantum fluctuations in Sec\ref{5}. The paper is concluded in Sec.\ref{6}.

\section{Building up the models: Calculation of interaction parameters}\label{2}

This section describes how to build-up the  models with a pair of interacting atoms in a DW potential. We consider a 3D trapping potential of the form 
\begin{eqnarray}
V_{trap}(r)= \frac{1}{2}m{\omega_\rho}^2{\rho}^2+\frac{1}{2}\lambda^2(z^2-\eta^2)^2
\label{eq1}
\end{eqnarray}
where $\rho^2 = x^2 + y^2$,  $\omega_\rho$ is radial trapping frequency,  $z = {\pm\eta}$ are the two minimum points where the trapping potential along the $z$-axis vanishes and the barrier height of the DW is $V_0=\frac{1}{2}\lambda^2\eta^4$. Here we have assumed that the 3D trap has  harmonic oscillations along radial directions ($x$- and $y$-axes) and a DW  along $z$-axis. The DW configuration along the $z$-direction can be created by combining a repulsive Gaussian potential with a 1D harmonic trap, and then expanding the Gaussian potential upto the 4th order in $z$ one can obtain the above form of the DW trap. If the barrier height $V_0$ is very
large compared to the ground-state energy, each well will behave like an  almost independent harmonic oscillator. Under this harmonic approximation near $z= \pm \eta$, we get the harmonic frequency  $\omega_z=\frac{2\lambda\eta}{\sqrt{m}}$. 
We assume that the temperature is low enough so that the atoms occupy only the ground state of the radial harmonic potentials. 
The aspect ratio between the radial and axial trap size is defined by  $ \zeta = \sqrt{\omega_z/\omega_\rho}$. 

By integrating over the radial harmonic oscillator states, one can obtain an effective 1D Hamiltonian for the system. We solve for single-particle 1D eigenfunctions and eigenvalues numerically using the method of discrete variable representation (DVR). For symmetric DW, the lowest eigenstate $\psi_s(z)$ is space-symmetric ($\psi_s(z) = \psi_s(-z)$) and the other quasi-degenerate state $\psi_a(z)$ is antisymmetric 
($\psi_a (z) = - \psi_a(-z)$). One can form  two-mode basis states $\psi_{\pm}(z) = [ \psi_s(z) \pm \psi_a(z)]/\sqrt{2}$. Under tight-binding approximation ($\hbar \omega_z <\!< V_0$), $\psi_{\pm}(z)$ are substantially localized either on the left or right well of the DW. Let us rename $\psi_{l} = \psi_+(z)$  and $\psi_{r} = \psi_-(z)$ as the left and the right localized states, respectively. We then obtain the tunnel coupling $J$ by calculating the matrix element $-\hbar J = \int d z \psi_l(z) H_1 \psi_r(z)$, where $H_1 = p_z^2/2m + V_{dw}(z)$ is the 1D single-particle Hamiltonian with $V_{dw}(z) =  \frac{1}{2}\lambda^2(z^2-\eta^2)^2$ being the DW potential and $m$ being the mass of the particle. 

In terms of the site-specific (left-right) basis functions, there are in general four coefficients of interaction
\begin{equation} \tag{2.A}
U_{jj}  = \int\int |\Phi_j ({\bf r_1})|^2 V_{int}(|{\bf r_1} - {\bf r_2}|) |\Phi_j ({\bf r_2})|^2 d{\bf r_1}d{\bf r_2} \nonumber
\label{eq2A}
\end{equation}
\begin{equation} \tag{2.B}
U_{i} = \int \int |\Phi_j ({\bf r_1})|^2 V_{int}(|{\bf r_1} - {\bf r_2}|) |\Phi_k ({\bf r_2})|^2 d{\bf r_1}d{\bf r_2}, \hspace{0.5 cm} j \ne k \nonumber
\label{eq2B}
\end{equation}
\begin{equation} \tag{2.C}
I = \int \int \Phi_j^{\dagger} ({\bf r_1}) \Phi_j^{\dagger} ({\bf r_2}) V_{int}(|{\bf r_1} - {\bf r_2}|) \Phi_j ({\bf r_1}) \Phi_k ({\bf r_2}) d{\bf r_1}d{\bf r_2} \hspace{0.5 cm} j \ne k \nonumber
\label{eq2C}
\end{equation}
\begin{equation} \tag{2.D}
K = \int \int \Phi_j^{\dagger} ({\bf r_1}) \Phi_k^{\dagger} ({\bf r_2}) V_{int}(|{\bf r_1} - {\bf r_2}|) \Phi_j ({\bf r_2}) \Phi_k ({\bf r_1}) d{\bf r_1}d{\bf r_2}, 
\hspace{0.5 cm} j \ne k
\label{eq2D}
\end{equation}
where `$j$' and `$k$' stand for the site index `$l$' (left) and `$r$' (right), $\Phi_j ({\bf r_j}) = \phi_0(\rho_j) \psi_j(z_j)$ with $\phi_0(\rho_j)$ is the ground state of 2D harmonic oscillator wave function of $j$th particle, $V_{int}(|{\bf r_1} - {\bf r_2}|)$ denotes the interaction potential between the two particles `1' and `2'. Here $U_{i}$, $I$ and $K$ are inter-site, partial exchange and total exchange interaction terms, respectively.  $U_{ll}$ and $U_{rr}$ are the left and right on-site interaction terms, respectively. For a symmetric DW potential, we have $U_{ll}=U_{rr} = U$. These four types of interaction terms are schematically shown in Fig.\ref{Figure 1.}.

\begin{figure}[h]
        \centering
        \includegraphics[width=4.2in, height=2.4in]{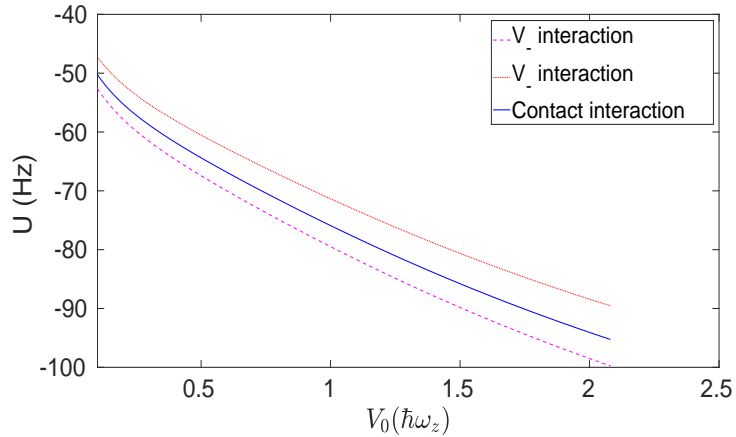}
        \caption{Comparison between on-site contact and finite-range Jost-Kohn interaction terms $U$ for negative $a_s$ as a function of the height of the potential barrier ($V_0$) of the DW. Note that the full six dimensional wave functions of the interacting atom-pair are used to calculate $U$ (see the text). The  Jost-Kohn $U$ is calculated for two  ranges $r_0= r_{\rm vdW}$ ( red dotted) and $r_0=r_{\rm vdW}/100$ (pink dashed), with $a_s=-9.54$ nm and $r_{\rm vdW}=1.66$ nm which corresponds to that for the van der Waals' potential between a pair of $^6$Li atoms.}
        \label{Figure 2.}
    \end{figure}

It is important to choose an appropriate model interaction potential to represent the potential $V_{int}(|{\bf r_1} - {\bf r_2}|)$. Usually, for a standard Hubbard model for ultracold atoms, $V_{int}(r)$ is replaced by the zero-range delta-type pseudo-potential which is found to be applicable when the $s$-wave scattering length $a_s$ is much smaller than the length scale of the trap under harmonic approximation \cite{Schneider}. Furthermore, this delta-potential approximation breaks down when the effective range of interaction is finite or large as in the case of magnetic Feshbach resonances \cite{effrange:recent}, particularly when the width of the resonance is very narrow 
\cite{Schneider,EPJ:2016}. To overcome these limitations of the contact potentials, we here resort to model finite-range interaction potentials  of Jost and Kohn \cite{JostKhon}  because these potentials hold good for a wide range of scattering lengths and arbitrary range. The usefulness of Jost-Kohn potentials to model atom-atom interactions has been discussed elsewhere \cite{our_arxiv_pap}. However, for the sake of completeness,  we briefly describe these potentials in Appendix \ref{appendix-A}. Here, as a consistency check, we verify whether these potentials can reproduce, at least qualitatively, the results of contact interaction under the appropriate limiting or physical conditions.  For this purpose, we numerically solve for the full six dimensional wave function of a pair of atoms interacting with negative $a_s$ in a 3D trap for which we assume that the radial harmonic trapping frequency $\omega_{\rho}$ is equal to the approximate axial trapping frequency $\omega_z$ calculated under harmonic approximation of the DW trap. We also calculate the same for contact potential. We then use these wave functions to calculate the on-site interaction parameter $U$. In Fig.\ref{Figure 2.}, we plot these results as a function of the barrier height $V_0$ by changing the parameter $\lambda$ keeping the minimum positions of the wells $\pm \eta$ fixed. We choose realistic parameters for these plots, considering a pair of $^6$Li atoms interacting with a small negative scattering length $a_s= -9.54 $ nm and the effective range $r_0$ being equal to the characteristic  van der Waal's length scale $r_{\rm vdW}=1.66$ nm \cite{chin:rmp:2010}. From this figure, we notice that, under harmonic or tight-binding approximation ($V_0 > \hbar \omega_z$), both the results for contact and Jost-Kohn potentials vary linearly with $V_0$ and quantitatively they do not deviate much. This figure also shows that as the range is decreased by 2 order of magnitude, $U$ decreases by about less than 10 percent.  These results signify that for small scattering length and effective range, 
and in almost isotropic 3D trapping situation, the Jost-Kohn potential and the contact potential yield almost similar Hubbard interaction parameters. It has also been shown earlier that the results of contact interactions are almost reproducible by Jost-Kohn potentials only in case of free-space or isotropic trap provided scattering length and the range are sufficiently small \cite{partha_physscrpt}. In this context, it is worth mentioning that, the  validity of the exact solutions \cite{busch:foundphys} of the problem of two cold atoms  in a harmonic oscillator interacting via the regularized contact potential is restricted to the sufficiently weak isotropic trap for which the trap size should be much larger than $|a_s|$  \cite{Tiesinga_2000}. The regularized contact potential with an energy-dependent T-matrix element can also serve as a model potential for calculating effective range effects \cite{Bolda_02,PHYSICAL REVIEW A 83 030701(R)}. The self-consistent method using this potential has been found to be inadequate to estimate correctly the Hubbard interaction parameters even if $a_s$ is much smaller than the length 
scale of the trap \cite{schneider:pra:2009}.

 Having established the equivalence between the contact and the Jost-Kohn potential $V_{-}(r)$ of negative $a_s$ for an almost isotropic trap provided both $a_s$ and $r_0$ are small enough, we now calculate $U$ using Jost-Kohn potentials of both positive and negative $a_s$ for  a quasi-1D DW trap for which $\omega_{\rho}$ is assumed to be much larger that $\omega_z$. Here we make use of the usual Hubbard approximation: Wave functions of non-interacting particles in the trap under tight-binding approximation are used to calculate the interaction matrix elements. The method of calculation of $U$ under Hubbard approximation is briefly discussed in Appendix-\ref{appendix-B}. We compare these matrix elements with those obtained using the pair wave functions that take into account the effect of interaction potentials. The results are displayed in Fig.\ref{Figure 3.}. As discussed in Appendix-\ref{appendix-A}, the Jost-Kohn potential $V_{+}(r)$ for positive scattering length depends on a third parameter $\kappa$ apart from $r_0$ and $a_s$. In the limit $\kappa \rightarrow \infty$ ($\kappa r_0 >\!> 1$), the scattering solution of $V_{+}(r)$ 
yields an effective range expansion with range $r_0$ ($>0$). However, in the limit $\kappa \rightarrow 0  $ ($\kappa r_0 <\!< 1$), the effective range expansion gets modified yielding a modified range which may become negative and quite large \cite{our_arxiv_pap}. For the plots in Fig.\ref{Figure 3.}, we use the former condition, that is, $\kappa r_0 >\!> 1$. From this figure, we observe that, though both the results obtained under the Hubbard approximation and beyond this approximation provide qualitatively similar results if $r_0$ is small, the results with Hubbard approximation are underestimated by about 1/3 compared to those corresponding to the solutions of interaction Hamiltonian of two trapped atoms. We further notice that $U$ saturate at large $|a_s|$, the larger the value of $|r_0|$, the larger is the value of $|a_s|$ at which $U$ saturates. For small $|a_s|$ regime, $U$ varies almost linearly with $|a_s|$ as in the case of contact interaction. In the other condition $\kappa r_0 <\!< 1$ for positive $a_s$, $U$ 
varies highly nonlinearly as shown elsewhere \cite{Abhik_paper}. So far we have presented the results on $U$ only. The other three interaction parameters $U_i$, $K$ and  $I$ are usually smaller than $U$ by two orders of magnitude for small $r_0$, otherwise they show the similar dependence on $|a_s|$ as in the case of $U$. However,  condition the $\kappa r_0 <\!< 1$ which may apply to a narrow Feshbach resonance, $U$ may become zero or negative near the resonance while the the other three interaction parameters remaining finite \cite{Abhik_paper}.   
 
 So far we have discussed calculation of interaction parameters for Hubbard model using  finite-range and contact interactions. Before ending this section, we would like to briefly address a related  
 question: How do the purely long-range interactions which vary as inverse power law of the interatomic separation,  such as magnetic dipole-dipole interaction of dipolar atoms can affect the results of Hubbard models?  We have calculated $U$ and $U_i$ for dipolar Chromium atoms when the dipole moments are oriented parallel to each other. The magnetic dipole moment of Cr is $\approx6\mu_B$ ($\mu_B$ is bohr magneton). The  calculated dipolar on-site interaction  $U$ turns out to be about  1.3 kHz and inter-site interaction $U_i$  is  about 16.8 Hz. These results indicate that it is possible to  include the long-range DDI of dipolar systems within the framework of extended Hubbard models with all four interaction terms.

    \begin{figure}[h]
    \centering
    \begin{minipage}{0.48\textwidth}
        \centering
        \includegraphics[width=3.2in, height=2.0in]{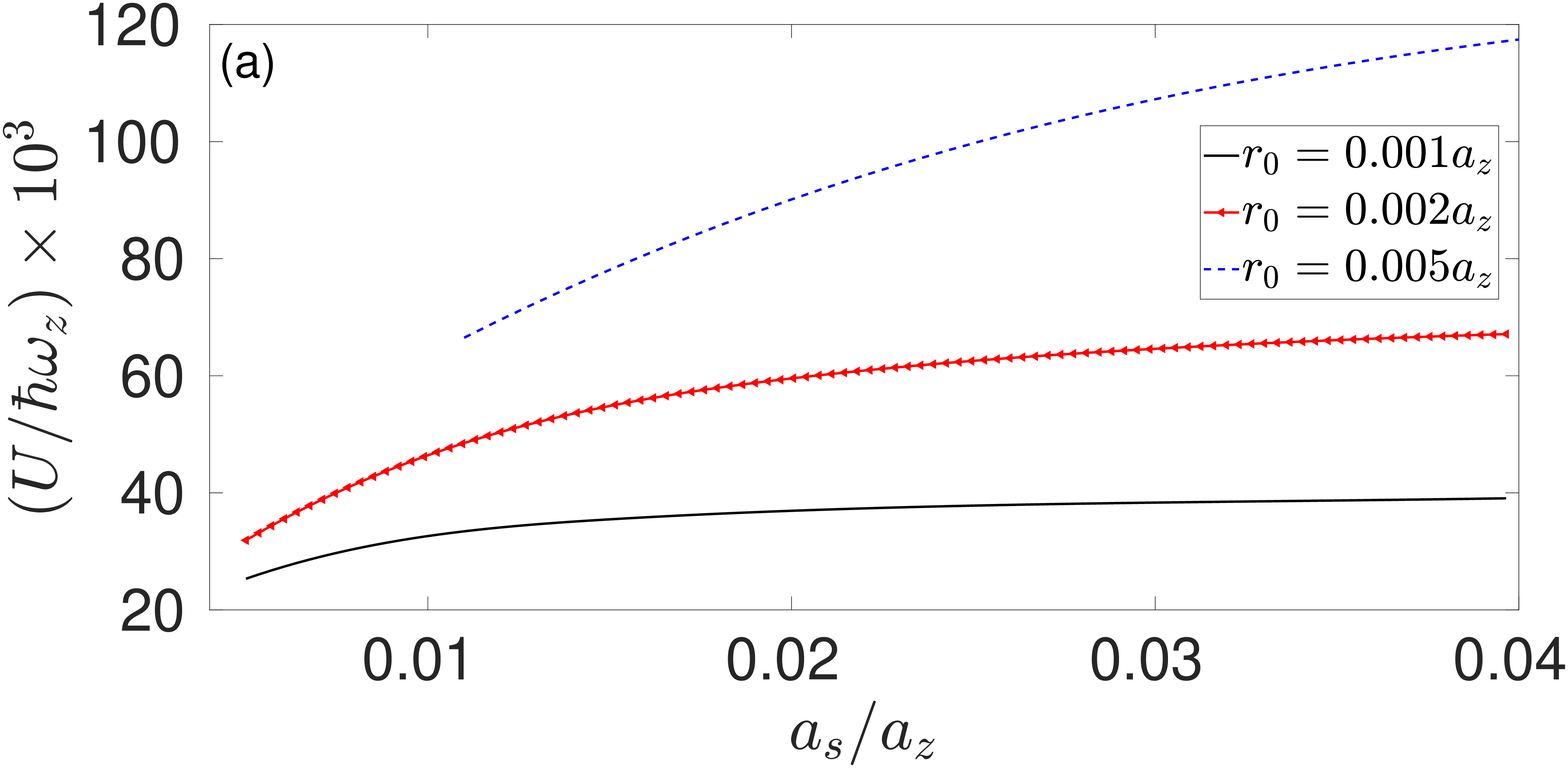}
    \end{minipage}
        \begin{minipage}{0.48\textwidth}
        \centering
        \includegraphics[width=3.2in, height=2.0in]{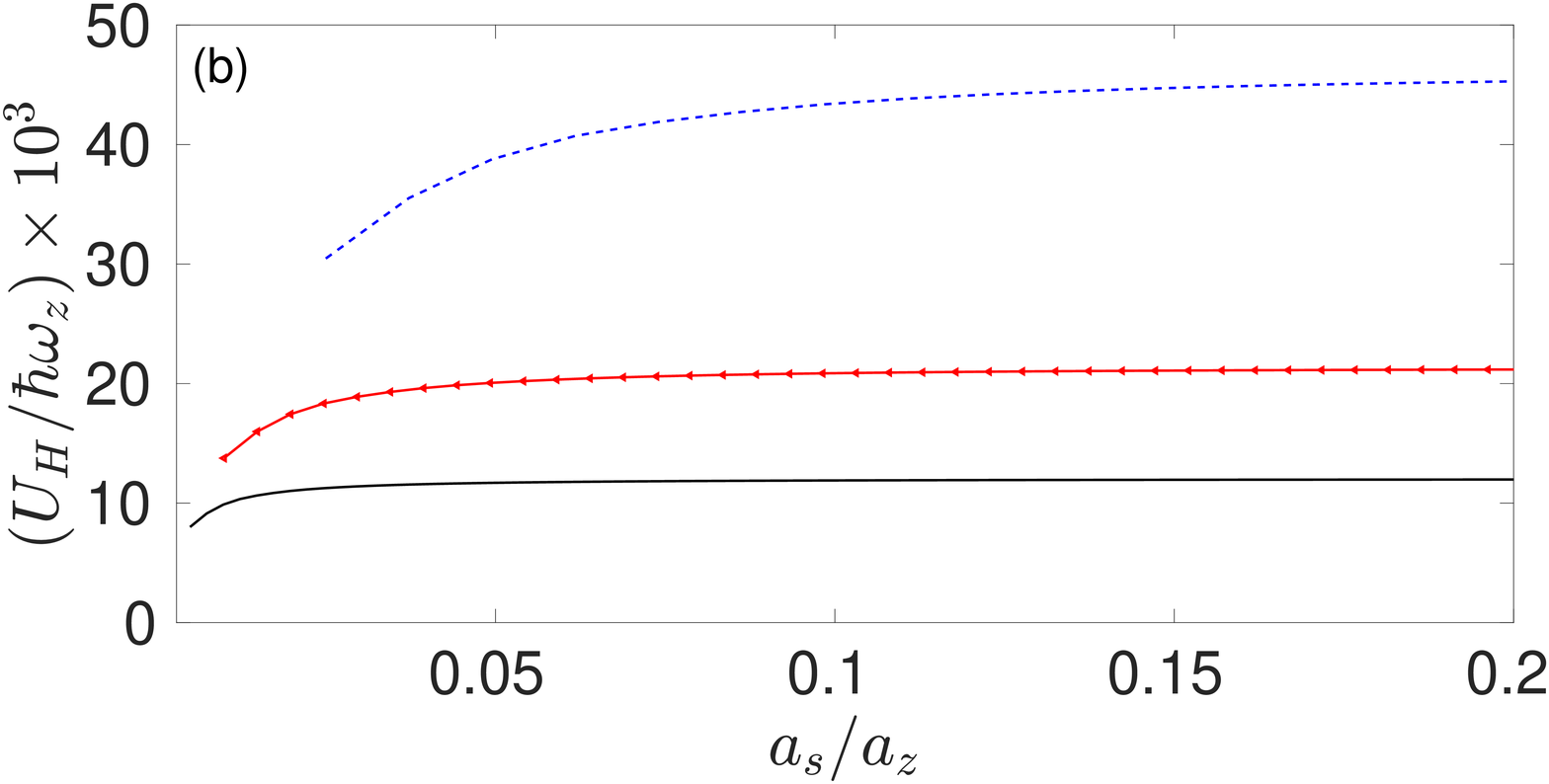}
    \end{minipage}
    \begin{minipage}{.48\textwidth}
        \centering
        \includegraphics[width=3.2in, height=2.0in]{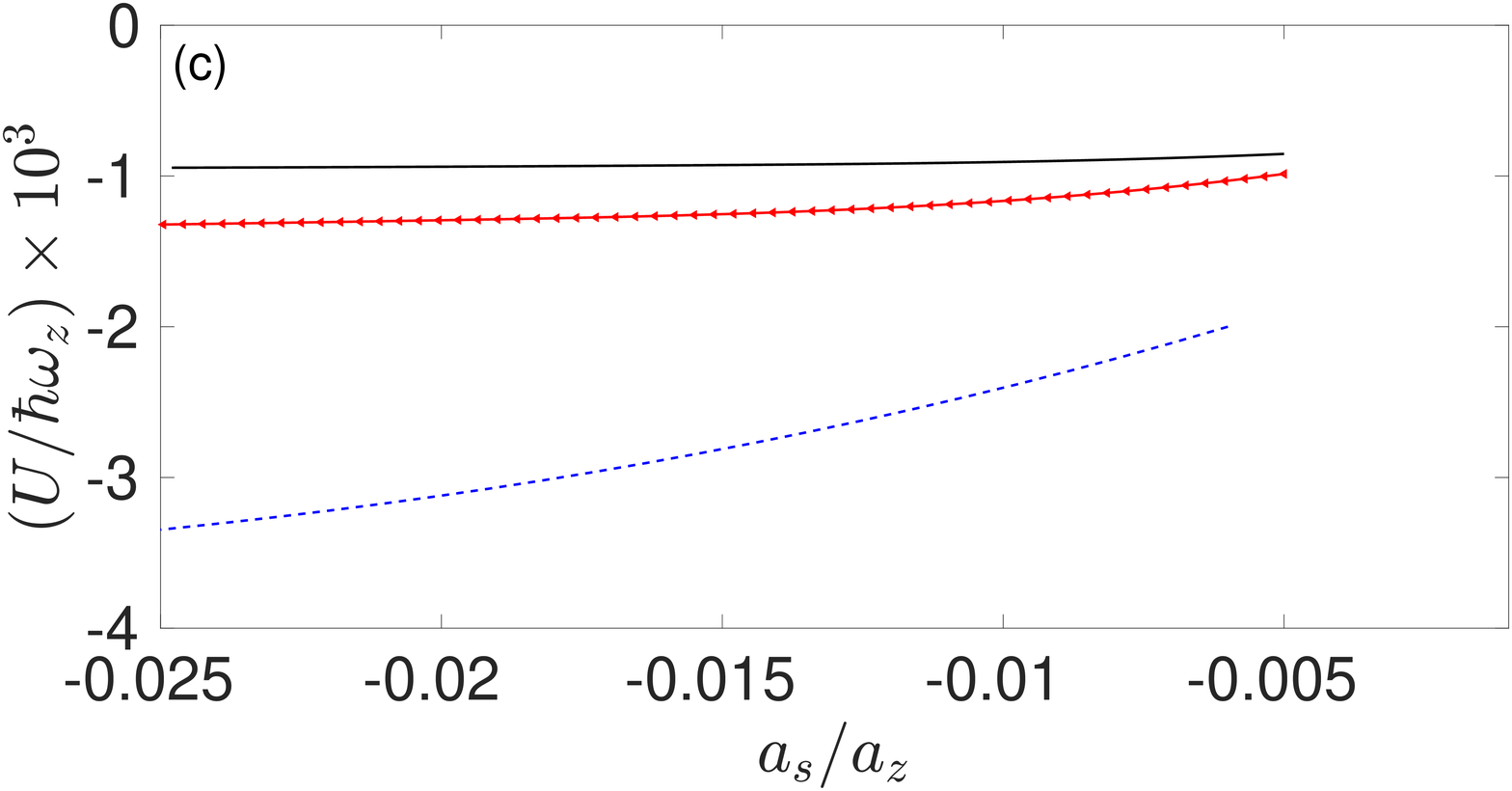}
    \end{minipage}
        \begin{minipage}{0.48\textwidth}
        \centering
        \includegraphics[width=3.2in, height=2.0in]{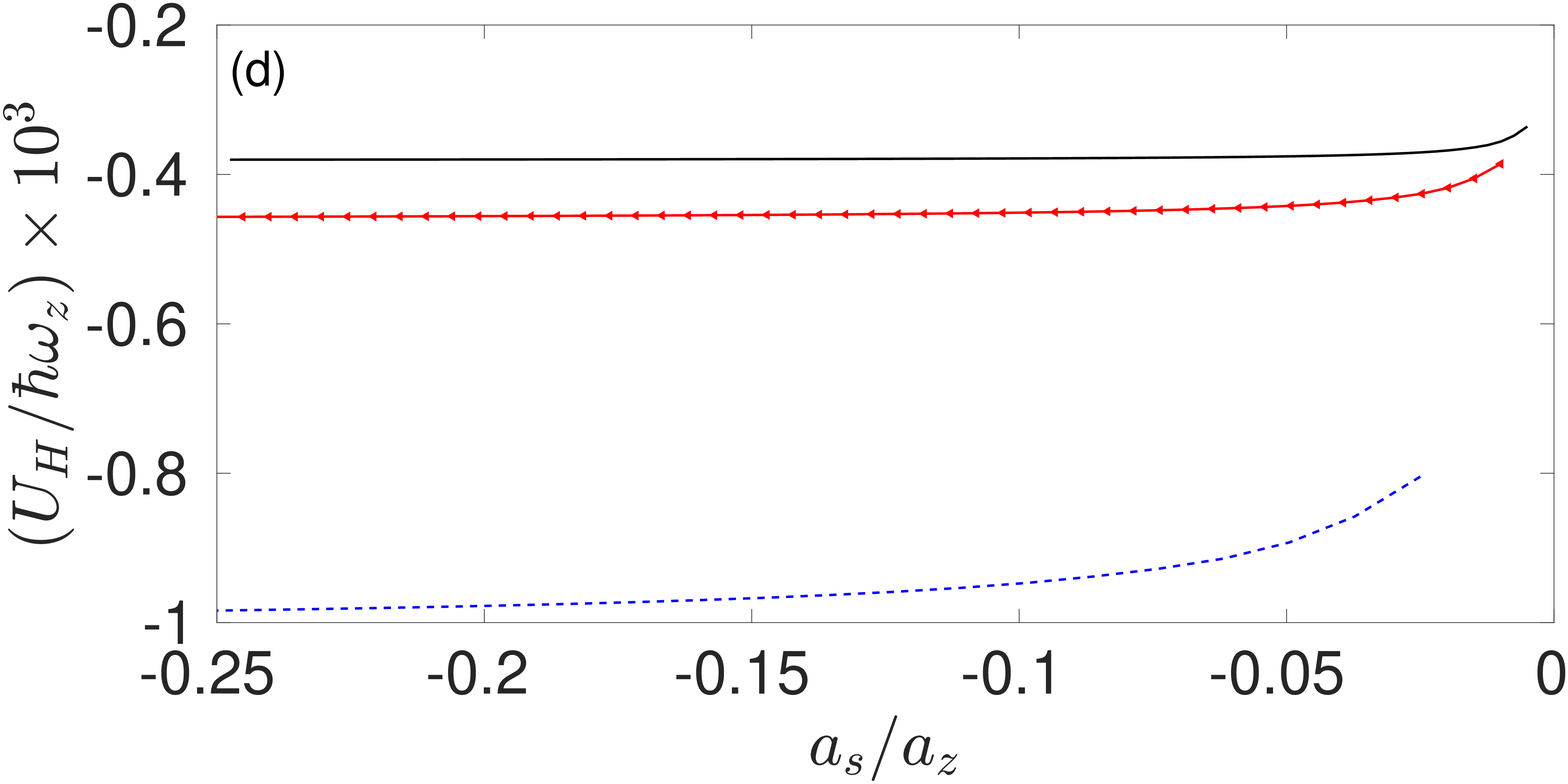}
    \end{minipage}
 \caption{\small Variation of $U$ (in unit of $\hbar\omega_z$) as a function of positive (a,c) and negative (b,d) $a_s$ (in unit of $a_z$) for three different values of  $r_0$. For (a) and (c) $U$ is calculated by solving the interaction Hamiltonian of two trapped atoms while for (b) and (d) $U$ is calculated under tight-binding Hubbard approximation. Here  $\omega_z/\omega_\rho=0.01$ and $\kappa a_z = 50000$. }
 \label{Figure 3.}
\end{figure}

We next present analytical and numerical solutions for both fermionic and bosonic two-site Hubbard models.

 \section{Two-site fermionic and bosonic Hubbard models}\label{3}
 
\subsection{A pair of two-component fermions}\label{3.1}
Let us consider a pair of two-component fermions in a DW potential under tight-binding approximation. Let the two components be denoted by the spin states $|\uparrow\rangle$ ans $|\downarrow\rangle$. 
 The Hamiltonian of the system in localized basis is
\begin{widetext}
\begin{eqnarray}
\hat{H_F} &=&   
-J\sum_{\sigma,\sigma'}\sum_{\alpha,\beta}\hat a^\dagger_{\alpha\sigma}\hat a_{\beta\sigma'}+ U\sum_{\sigma,\sigma'}\sum_{\alpha}\hat a^\dagger_{\alpha\sigma}\hat a^\dagger_{\alpha\sigma'}\hat a_{\alpha\sigma}\hat a_{\alpha\sigma'}\nonumber\\ &+& U_i\sum_{\sigma,\sigma'}\sum_{\alpha,\beta}\hat a^\dagger_{\alpha\sigma}\hat a^\dagger_{\beta\sigma'}\hat a_{\alpha\sigma}\hat a_{\beta\sigma'} + I\sum_{\sigma,\sigma'}\sum_{\alpha,\beta}(\hat a^\dagger_{\alpha\sigma}\hat a^\dagger_{\alpha\sigma'}\hat a_{\alpha\sigma}\hat a_{\beta\sigma'}+\hat a^\dagger_{\alpha\sigma}\hat a^\dagger_{\beta\sigma'}\hat a_{\alpha\sigma}\hat a_{\alpha\sigma'})\nonumber\\ &+& K\sum_{\sigma,\sigma'}\sum_{\alpha,\beta} (\hat a^\dagger_{\alpha\sigma}\hat a^\dagger_{\beta\sigma'}\hat a_{\beta\sigma}\hat a_{\alpha\sigma'}+\hat a^\dagger_{\alpha\sigma}\hat a^\dagger_{\alpha\sigma'}\hat a_{\beta\sigma}\hat a_{\beta\sigma'}) 
\end{eqnarray}
\end{widetext}
where $\hat{a}_{s\sigma} (\hat{a}^{\dagger}_{s\sigma})$ represents annihilation (creation) operator of a fermion in site $s$ ($\equiv l,r$) and spin state $\sigma$ ($\equiv \uparrow, \downarrow$).

Including spin and space variables, the single-particle state of the $i$th particle with spin $\sigma$ in the $j$th well may be denoted by $\psi_j(r_i \sigma ) \equiv \psi_j(r_i) \mid \sigma_i \rangle$, where $\sigma \equiv \uparrow,\downarrow$.  For a pair of two-component fermions, we have 4 uncoupled product basis  
\begin{eqnarray}
 |\uparrow\downarrow,0 \rangle &=& \frac{1}{\sqrt{2}} \left[|L_1\uparrow,L_2\downarrow \rangle - |L_1\downarrow,L_2\uparrow\rangle\right] \\
 |\uparrow,\downarrow\rangle &=& \frac{1}{\sqrt{2}} \left [|L_1\uparrow,R_2\downarrow \rangle - |R_1\downarrow,L_2\uparrow\rangle \right ] \\
 |\downarrow,\uparrow\rangle &=& \frac{1}{\sqrt{2}}\left[|L_1\downarrow,R_2\uparrow\rangle -|R_1\uparrow,L_2\downarrow\rangle\right] \\
 |0,\uparrow\downarrow\rangle &=&\frac{1}{\sqrt{2}} \left [|R_1\uparrow,R_2\downarrow\rangle - |R_1\downarrow,R_2\uparrow\rangle \right ]
\end{eqnarray}
where 
\bea 
|L_i\sigma,L_j\sigma'\rangle &=& \psi_l(r_i)|\sigma\rangle \otimes \psi_l(r_j)|\sigma'\rangle\\
|L_i\sigma,R_j\sigma'\rangle &=& \psi_l(r_i)|\sigma\rangle \otimes \psi_r(r_j)|\sigma'\rangle
\eea
One can rewrite these states in terms of spin singlet and spin triplet state of two particles (with $s=1$ and $m_s=0$)
\bea 
\label{eq:s}
\mid s \rangle = \frac{1}{\sqrt{2}} \left [ \mid \uparrow_1 \downarrow_2\rangle - \mid \downarrow_1 \uparrow_2\rangle \right ]\\
\mid t \rangle = \frac{1}{\sqrt{2}} \left [ \mid \uparrow_1 \downarrow_2\rangle + \mid \downarrow_1 \uparrow_2\rangle \right ]
\label{eq:t}
\eea
Thus, one can notice that for a pair of spin-$1/2$ particles in the lowest band of a DW trap, the particles is in spin-singlet when they occupy the same site while they are in the superposition of singlet and triplet when they reside in two different sites. 

\begin{figure}[h]
\centering
    \centering
    \begin{minipage}{0.48\textwidth}
        \centering
        \includegraphics[width=3.4in, height=2.2in]{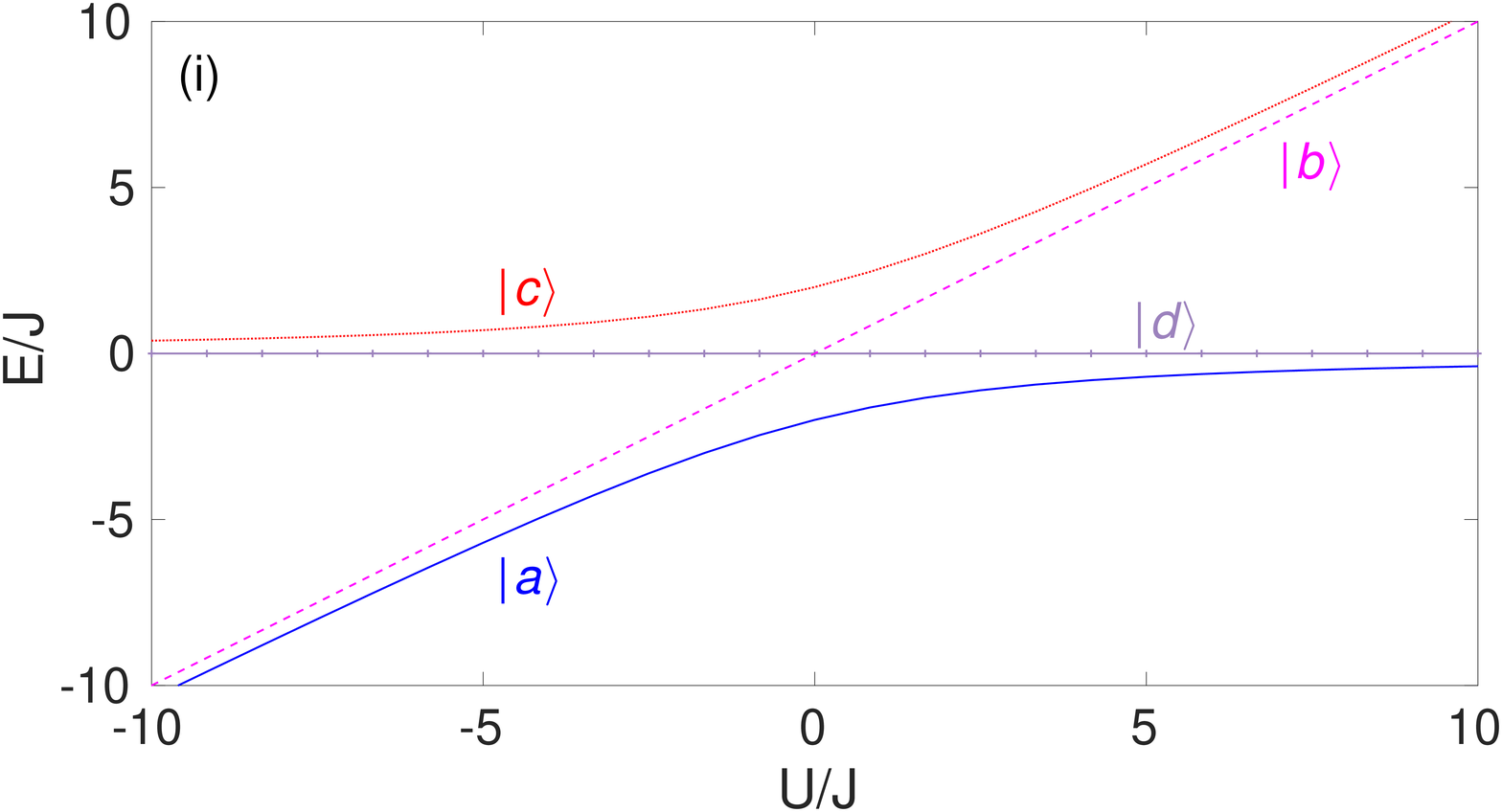}
    \end{minipage}
    \begin{minipage}{0.48\textwidth}
        \centering
        \includegraphics[width=3.4in, height=2.2in]{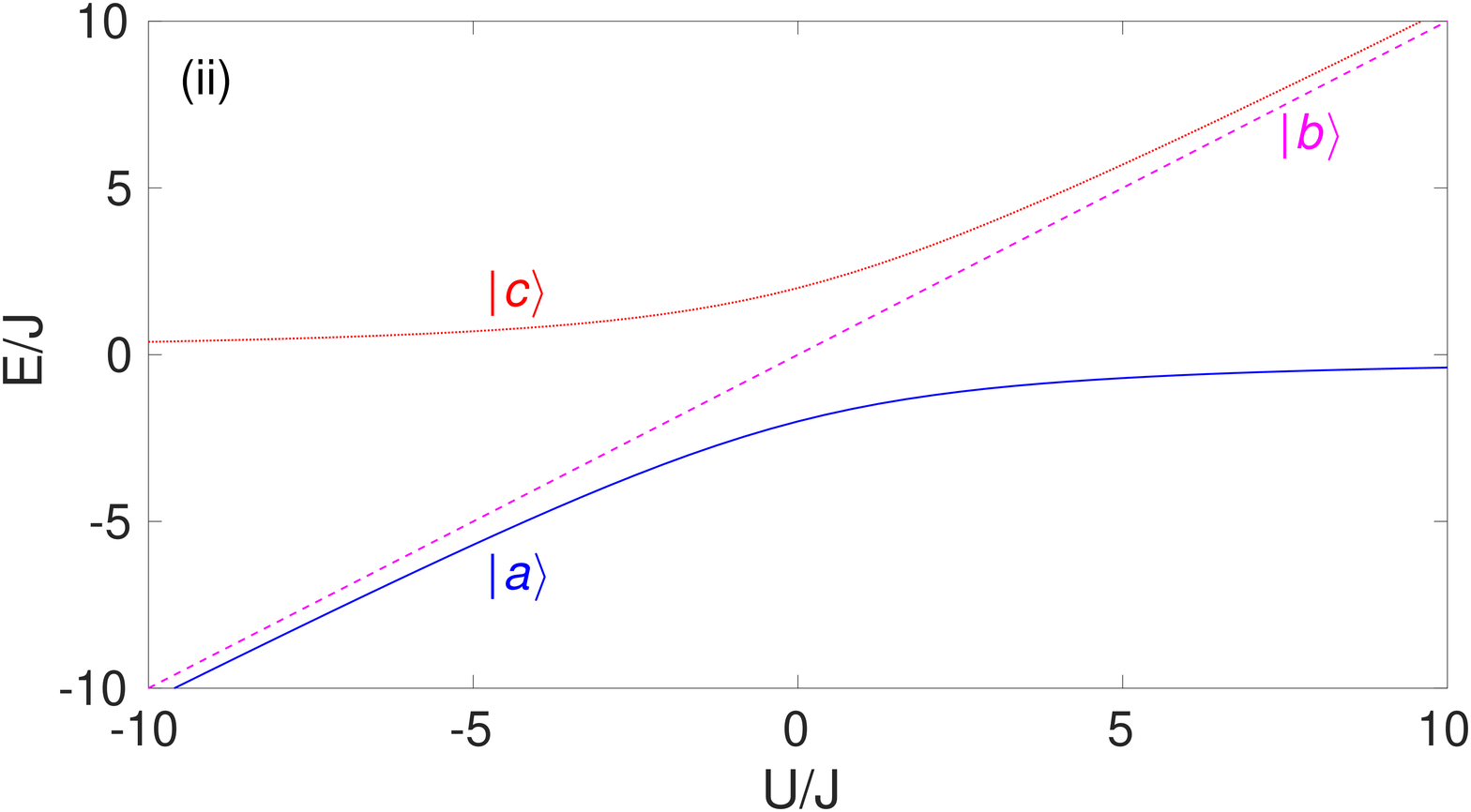}
    \end{minipage}
 \caption{\small The variation of dimensionless energies $E/J$ of the eigenstates for a pair of fermions (i) and bosons (ii) as a function of dimensionless on-site interaction $U/J$.}
 \label{Figure 4.}
\end{figure}

Using these bases, the Hamiltonian for a symmetrical DW can be written in a matrix form 
 \begin{eqnarray}
  H_F=
   \left[ {\begin{array}{cccc}
   U & -J_{-} & -J_{-} & K \\
   -J_{-} & U_i & K & -J_{-} \\
   -J_{-} & K & U & -J_{-} \\
   K & -J_{-} & -J_{-} & U_i \\
  \end{array} } \right]
  \label{eq8}
\end{eqnarray}
where $J_{-}=J-I$. Let the four eigenvalues of the Hamiltonian be denoted by $E_a$, $E_b$, $E_c$ and $E_d$, with corresponding eigen functions $|a\rangle$, $|b\rangle$, $|c \rangle$ and $|d\rangle$, respectively. Let $\bar{U} = U + U_i+2K$, $U_{\pm} = (U\pm U_i)$ and $\Omega = \sqrt{U_{-}^2 + 16{J_{-}}^2}$. Explicitly, the eigenvalues are given by 
\begin{eqnarray}
\label{eqEa}
E_a = \frac{1}{2} \left ( \bar{U}-\Omega \right ), \\ 
E_c = \frac{1}{2} \left ( \bar{U}+\Omega \right ),
\label{eqEc}
\end{eqnarray}
$E_b = U-K$,  and $E_d = U_i-K$ (Fig.\ref{Figure 4.}). Clearly, $E_a$ and $E_c$ are the lowest and highest energy eigenvalues. The corresponding eigen functions are given by 
\begin{eqnarray}
\label{eq:eigena}
|a\rangle &=& \frac{4J_{-}}{\sqrt{16J_{-}^2+\left(U_{-}+\Omega\right)^2}}\left(|+\rangle + \frac{U_{-}+\Omega}{4J_{-}}|1\rangle \right) \\
|b\rangle &=& |-\rangle
\label{eq:eigenb}
\end{eqnarray}
\begin{eqnarray}
\label{eq:eigenc}
|c\rangle &=& \frac{4J_{-}}{\sqrt{16J_{-}^2+\left(U_{-}-\Omega\right)^2}}\left(|+\rangle + \frac{U_{-}-\Omega}{4J_{-}}|1\rangle \right) \\
|d\rangle &=& |0\rangle 
\label{eq:eigend}
\end{eqnarray}
where 
\bea 
|\pm \rangle &=& 1/\sqrt{2} \left (|\uparrow\downarrow,0\rangle \pm |0,\uparrow\downarrow\rangle \right )\\
|0\rangle &=& 1/\sqrt{2} \left (|\uparrow,\downarrow\rangle - |\downarrow,\uparrow\rangle \right )\\
|1\rangle &=& 1/\sqrt{2} \left (|\uparrow,\downarrow\rangle + |\downarrow,\uparrow\rangle \right )
\eea 

We next calculate the dynamical evolution of the system from an initial state which is not an eigen state  of the system.  Let the time-dependent wave function be expressed as 
$|\psi_F \rangle = c_0(t)|\uparrow\downarrow,0\rangle + c_1(t)|\uparrow,\downarrow\rangle + c_2(t)|\downarrow,\uparrow\rangle + c_3(t)|0,\uparrow\downarrow\rangle  $
 where $c_i(t)$  $(i=0,1,2,3)$ are the probability amplitude for the respective state. For the initial condition: $c_0(0) = 1$, $c_1(0) = 0$, $c_2(0) =0$ and $c_3(0) = 0$ i.e., both particles are initially in the left site, we obtain 
\begin{eqnarray}
 c_0(t)&=& 
 \frac{1}{2}e^{-itW}+\frac{1}{2}e^{-\frac{it\bar{U}}{2}}\Big[\cos\frac{\Omega t}{2}-\frac{iU_-}{\Omega}\sin\frac{\Omega t}{2}\Big] \nonumber\\
 c_1(t)&=&
  \frac{2iJ_{-}}{\Omega}e^{-\frac{it\bar{U}}{2}}\sin\frac{\Omega t}{2} \nonumber\\
  c_2(t)&=&
 \frac{2iJ_{-}}{\Omega}e^{-\frac{it\bar{U}}{2}}\sin\frac{\Omega t}{2} \nonumber\\
 c_3(t)&=&
 -\frac{1}{2}e^{-itW}+\frac{1}{2}e^{-\frac{it\bar{U}}{2}}\Big[\cos\frac{\Omega t}{2}-\frac{iU_-}{\Omega}\sin\frac{\Omega t}{2}\Big] 
 \label{eq9}
\end{eqnarray}
 where, $W=U-K$. If the system is initialized in $(|\uparrow,\downarrow \rangle+|\downarrow,\uparrow \rangle)/\sqrt{2}$, that is,  for the initial condition: $c_0(0) = 0$, $c_1(0) = \frac{1}{\sqrt{2}} = c_2(0)$ and $c_3(0) = 0$,  we have  
\begin{eqnarray}
 c_0(t) &=& c_3(t)\nonumber\\
 &=& \frac{2\sqrt{2}iJ_{-}}{\Omega}e^{-\frac{it\bar{U}}{2}}\sin\frac{\Omega t}{2}\nonumber\\
 c_1(t) &=& c_2(t)\nonumber\\
 &=& \frac{1}{\sqrt{2}}e^{-\frac{it\bar{U}}{2}}\Big[\cos\frac{\Omega t}{2}+\frac{iU_-}{\Omega}\sin\frac{\Omega t}{2}\Big] 
 \label{eq10}
\end{eqnarray}

\subsection{A pair of spinless as well as a pair of two-component bosons}\label{3.2}
The Hamiltonian for a pair of spinless bosons in a DW trap in the localized basis
\begin{widetext}
\begin{eqnarray}
 \hat H_B &=&
 -J\sum_{\alpha,\beta}\hat a^\dagger_\alpha\hat a_\beta+\frac{U}{2}\sum_{\alpha}\hat a^{\dagger2}_\alpha\hat a^{2}_\alpha+\frac{U_i}{2}\sum_{\alpha,\beta}\hat a^\dagger_\alpha\hat a^\dagger_\beta\hat a_\beta\hat a_\alpha\nonumber\\ &+& I\sum_{\alpha,\beta}(\hat a^{\dagger2}_\alpha\hat a_\alpha\hat a_\beta+\hat a^\dagger_\alpha\hat a^\dagger_\beta\hat a^{2}_\alpha)+\frac{K}{2}\sum_{\alpha,\beta}(\hat a^{\dagger2}_\alpha\hat a^{2}_\beta+\hat a^\dagger_\alpha\hat a^\dagger_\beta\hat a_\beta\hat a_\alpha)
\end{eqnarray}
\end{widetext}
where the site indices $L$, $R$ are denoted by $\alpha$ and $\beta$ and $\alpha\neq\beta$.

For a pair of spinless bosonic system there are three fock basis states
\begin{eqnarray}
 |2,0\rangle &=& \psi_l(r_1)\psi_l(r_2) \nonumber\\ 
 |1,1\rangle &=& \frac{1}{\sqrt{2}}\Big(\psi_l(r_1)\psi_r(r_2) + \psi_r(r_1)\psi_l(r_2)\Big)\nonumber\\
 |0,2\rangle &=& \psi_r(r_1)\psi_r(r_2)\nonumber\\
\end{eqnarray}
where $\mid n,m\rangle$ implies a state containing $n$ and $m$ bosons in left and right well, respectively. The states are exactly same for single-component spin polarized bosonic system.

Now, let us consider a pair of two-component bosons, where two components refer to the two hyperfine spin states denoted by $\mid \uparrow \rangle \equiv \mid F, m_F \rangle$, $ \mid \downarrow\rangle \equiv \mid F, m_F' \rangle$ where $F$ is the hyperfine quantum number, $m_F$ and $m_F'$ are its two projections along the quantization axis.  The bases can be written as 
\begin{eqnarray}
 |\uparrow\downarrow,0 \rangle &=& \frac{1}{\sqrt{2}} \left[|L_1\uparrow,L_2\downarrow \rangle + |L_1\downarrow,L_2\uparrow\rangle\right] \\
 |\uparrow,\downarrow\rangle &=& \frac{1}{\sqrt{2}} \left [|L_1\uparrow,R_2\downarrow \rangle + |R_1\downarrow,L_2\uparrow\rangle \right ] \\
 |\downarrow,\uparrow\rangle &=& \frac{1}{\sqrt{2}}\left[|L_1\downarrow,R_2\uparrow\rangle +|R_1\uparrow,L_2\downarrow\rangle\right] \\
 |0,\uparrow\downarrow\rangle &=&\frac{1}{\sqrt{2}} \left [|R_1\uparrow,R_2\downarrow\rangle + |R_1\downarrow,R_2\uparrow\rangle \right ]
\end{eqnarray}
In contrast to the two-component fermion case, a pair of two-component bosons are always in spin-triplet-like state when both the bosons occupy the same site while the other two bases are the same as in the case of a pair of two-component fermions.  

The Hamiltonian in the case of a pair of spinless bosons can be expressed in the matrix form 
\begin{eqnarray}
  H_B=
  \left[ {\begin{array}{ccc}
   U & -\sqrt{2}J_{-} & K \\
   -\sqrt{2}J_{-} & U_i+K & -\sqrt{2}J_{-} \\
   K & -\sqrt{2}J_{-} & U \\
  \end{array} } \right]
  \label{eq11}
\end{eqnarray}
The three eigen functions for boson system can be readily obtained from the eigen functions $|a \rangle$, $|b \rangle$ and $|c \rangle$ of fermionic system  by replacing the bases  $ |\uparrow\downarrow,0\rangle \rightarrow |2,0\rangle $, 
 $ |0, \uparrow\downarrow \rangle \rightarrow |0, 2\rangle $ and 
   $ (|\uparrow, \downarrow \rangle + |\downarrow, \uparrow\rangle)/\sqrt{2} \rightarrow |1,1\rangle $. The corresponding eigenvalues $E_a$, $E_b$ and $E_c$ remain the same. As a result, many of the characteristics of bosonic system remain the same as that of the fermionic system, given the same input parameters.  

Let the time-dependent wave function of the spinless bosonic system be  
\begin{equation}
 |\psi_B\rangle = C_0(t)|2,0\rangle + C_1(t)|1,1\rangle + C_2(t)|0,2\rangle \nonumber
\end{equation}
The coefficients for initial condition $C_0(0) = 1$, $C_1(0) = 0$ and $C_2(0) = 0$ are given by 
\begin{eqnarray}
C_0(t)
&=& \frac{1}{2}e^{-itW}+\frac{1}{2}e^{-\frac{it\bar{U}}{2}}\Big[\cos\frac{\Omega t}{2} - \frac{iU_-}{\Omega}\sin\frac{\Omega t}{2}\Big]\nonumber\\
C_1(t)
&=& \frac{2\sqrt{2}iJ_{-}}{\Omega}e^{-\frac{it\bar{U}}{2}}\sin\frac{\Omega t}{2}\nonumber\\
C_2(t)
&=& -\frac{1}{2}e^{-itW}\nonumber\\&+&\frac{1}{2}e^{-\frac{it\bar{U}}{2}}\Big[\cos\frac{\Omega t}{2} - \frac{iU_-}{\Omega}\sin\frac{\Omega t}{2}\Big]
\label{eq12}
\end{eqnarray}

For another initial condition $C_0(0) = 0$, $C_1(0) = 1$, $C_2(0) = 0$, we have 
\begin{eqnarray}
 C_0(t) &=& C_2(t)  
 = \frac{2\sqrt{2}iJ_{-}}{\Omega}e^{-\frac{it\bar{U}}{2}}\sin\frac{\Omega t}{2}\nonumber\\
 C_1(t) 
 &=& e^{-\frac{it\bar{U}}{2}}\Big[\cos\frac{\Omega t}{2}+\frac{iU_-}{\Omega}\sin\frac{\Omega t}{2}\Big] 
 \label{eq13}
\end{eqnarray} 

From the above analytical calculations, we notice that, under otherwise similar physical conditions, the spatial part of the wave function of a pair of two-component fermions may be distinguished form that of spinless or two-component bosons by spin-specific measurements or the ``wave-function anatomy'' \cite{landman:pra:2018} or by analysing two-particle space or momentum correlation functions. Here we do not study such specific spin-dependent effects, instead we focus mainly on the average statistical features in the two cases. 

\subsection{Three two-component fermions vs. three spinless bosons}\label{3.3}

Let us consider three spinless bosons vs. three two-component fermions. For three bosons system there are four Fock basis: $|3,0\rangle, |2,1\rangle, |1,2\rangle$ and $|0,3\rangle$ and the Hamiltonian is,
 \begin{eqnarray}
 H_B=
 \left[ {\begin{array}{cccc}
          3U & -\sqrt{3}(J-2I) & \sqrt{3}K & 0 \\
          -\sqrt{3}(J-2I) & U+2U_i+2K & -2(J-2I) & \sqrt{3}K \\
          \sqrt{3}K & -2(J-2I) & U+2U_i+2K & -\sqrt{3}(J-2I) \\
          0 & \sqrt{3}K & -\sqrt{3}(J-2I) & 3U
         \end{array} }\right]\nonumber
\end{eqnarray}
On the other hand for three fermion system there are two possible combination of Fock basis. The wave function has either $|\uparrow\downarrow,\uparrow\rangle$ and $|\uparrow,\uparrow\downarrow\rangle$ or $|\uparrow\downarrow,\downarrow\rangle$ and $|\downarrow,\uparrow\downarrow\rangle$ with Hamiltonian
\begin{eqnarray}
  H_F = 
   \left[ {\begin{array}{cc}
   U+U_i & -J+2I \\
   -J+2I & U+U_i \\
  \end{array} } \right]
  \nonumber
\end{eqnarray}

From the Fock basis representation of the two Hamiltonians for three-particle systems,  it is evident that the two systems will have completly different static and  dynamical properties as there is no fermionic analog of the two bosonic states $|3,0\rangle$ and $|0,3\rangle$ unlike that in case of two-particle systems. Henceforth, we focus our attention to a comparative study between two-fermion vs. two-boson systems and do not consider  more than two particles any more.

\section{Static and dynamic properties}\label{4} 

For studying numerically the properties of the two models, we set the parameters  of the model DW potential of Eq.(1) such that,  under the harmonic approximation,  the trap frequency $\omega_z = 2\pi\times 1000$ Hz and  $J \approx 150$ Hz. 

\subsection{Eigenstructure}\label{4.1}

 In case of two fermions, for $U_{i} = 0$, the eigen energies reduce to those obtained in Ref. \cite{Murmann} where it is experimentally demonstrated that, once the system is prepared in the lowest state $|a \rangle$, a two-site counterpart of Mott insulator state can be realized by increasing the repulsive on-site interaction, while a two-site analog of charge density wave (CDW) state \cite{Ho} can be achieved by increasing the attractive on-site interaction. For $U>\,>J >> U_i$ one can find  $E_a \sim -4J^2/U$,  $E_c \sim  U + 4J^2/U$. Here $4J^2/U$ is the coupling of the second order tunneling matrix element \cite{Springer:assa} in the limit $U_i \rightarrow 0$ and $U \rightarrow \infty$. So, the two lowest eigenstates $|a\rangle$ and $|d\rangle$ can be coupled by second order tunneling process. Similarly, the transition between the excited states $|b\rangle$ and $|c\rangle$ is possible via second order process. From Eqs.(\ref{eqEa},\ref{eqEc}), it follows that, for $U_{-}>0$ and $U_{-} >\,>4J_-$, $|a \rangle \rightarrow  |1\rangle$, implying that when the on-site interaction is large positive the ground state of the system is the Mott-insulator where each site is occupied by a single particle. On the other hand, for $U_{-} <0$ and  $|U_{-}|>\,>4J_-$, $|a \rangle  \rightarrow |+ \rangle$ which is characterized by enhanced double occupancy, representing CDW phase. The CDW phase is dominated by second order pair tunneling that connects the state $|\uparrow\downarrow,0\rangle$ to $|0,\uparrow\downarrow\rangle$ via $|1\rangle$.  For $0< U_{-} >\,> 4J_-$,  the excited state $|c \rangle \rightarrow  |+ \rangle$ and  behaves like a CDW phase. An interesting question arises; What will happen if $4J <\, < U_i >\,> U$? In this limit,  Eqs.\ref{eqEa} and \ref{eqEc} yield $|a \rangle \rightarrow |+ \rangle$ and $|c \rangle \rightarrow  |1 \rangle$, that is for strong positive inter-site interactions, the ground state ($|a \rangle$) will be dominated by double occupancy and also probably CDW-like phase while the excited state ($|c \rangle$) will be like a Mott-Insulator with single occupancy. This happens because putting two atoms in the same well costs relatively less energy than to place them in two different wells. Next question is in which physical situation such a case may arise. Obviously this will not arise with zero-range interactions. However, for purely long-range interactions (with negligible short-range part) such as magnetic or electric dipole-dipole interactions this situation may occur.  

Setting $U_i =0$, we first plot the eigen energies  as a function of $U$ in Fig.\ref{Figure 4.} and reproduce the results reported in \cite{Murmann}. For the ground state $|a \rangle$ the double occupation probability decreases and single occupation probability increases as $U$ changes from negative to positive value keeping $U_i=0$. This means if the system is prepared in the ground eigenstate, by virtue of going from strong attractive to strong repulsive interaction regime ($|U|>\,> 4J$) , the system will undergo from CWD phase \cite{Ho} to  Mott-insulating phase \cite{Mott}.

\subsection{Occupation and tunneling}\label{4.2}

\begin{figure}[h]
\centering
   \begin{minipage}{.32\textwidth}
   \centering
    \includegraphics[width=2.2in, height=1.8in]{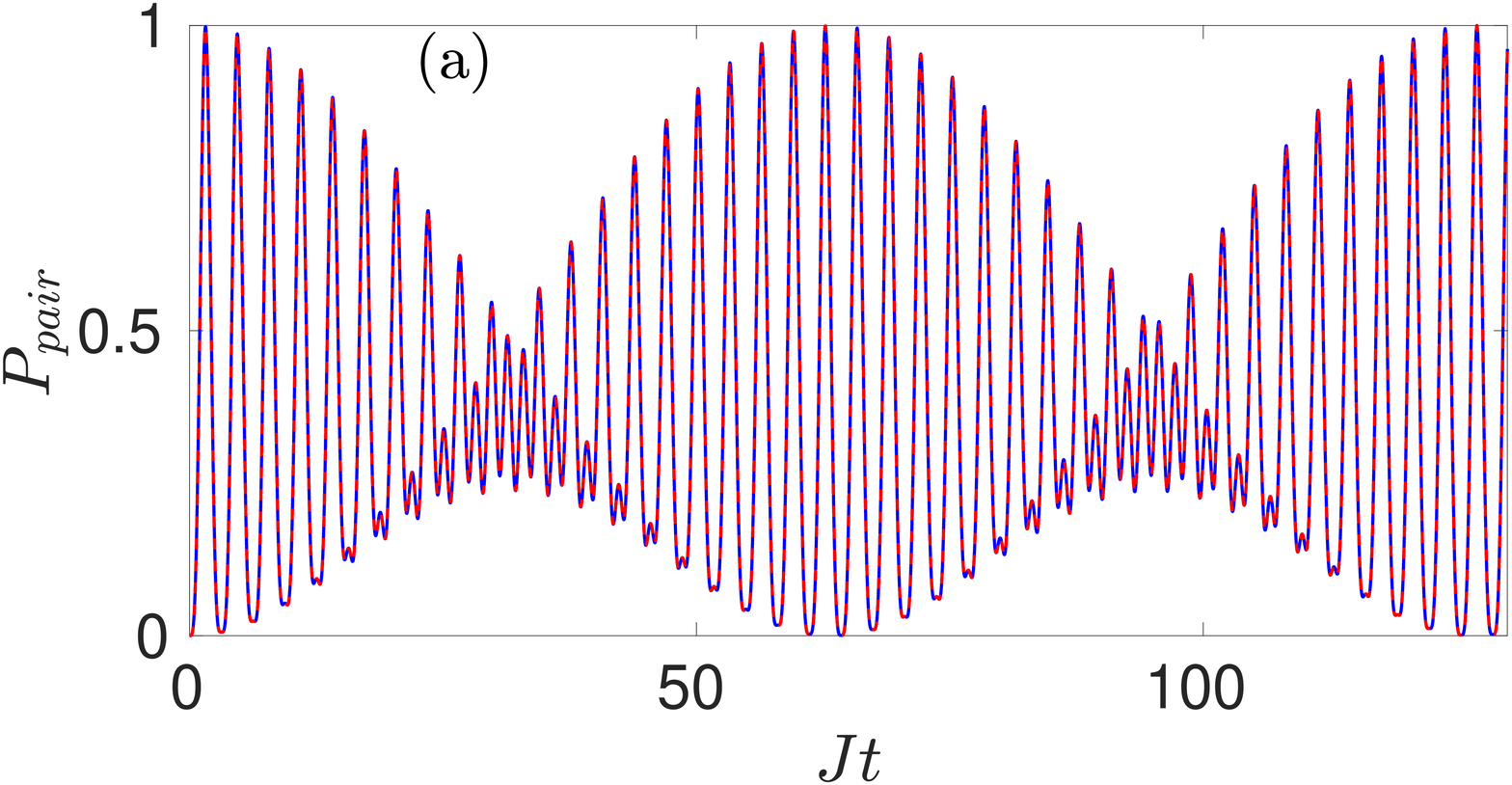} 
    \end{minipage}
    \begin{minipage}{.32\textwidth}
        \centering
        \includegraphics[width=2.2in, height=1.8in]{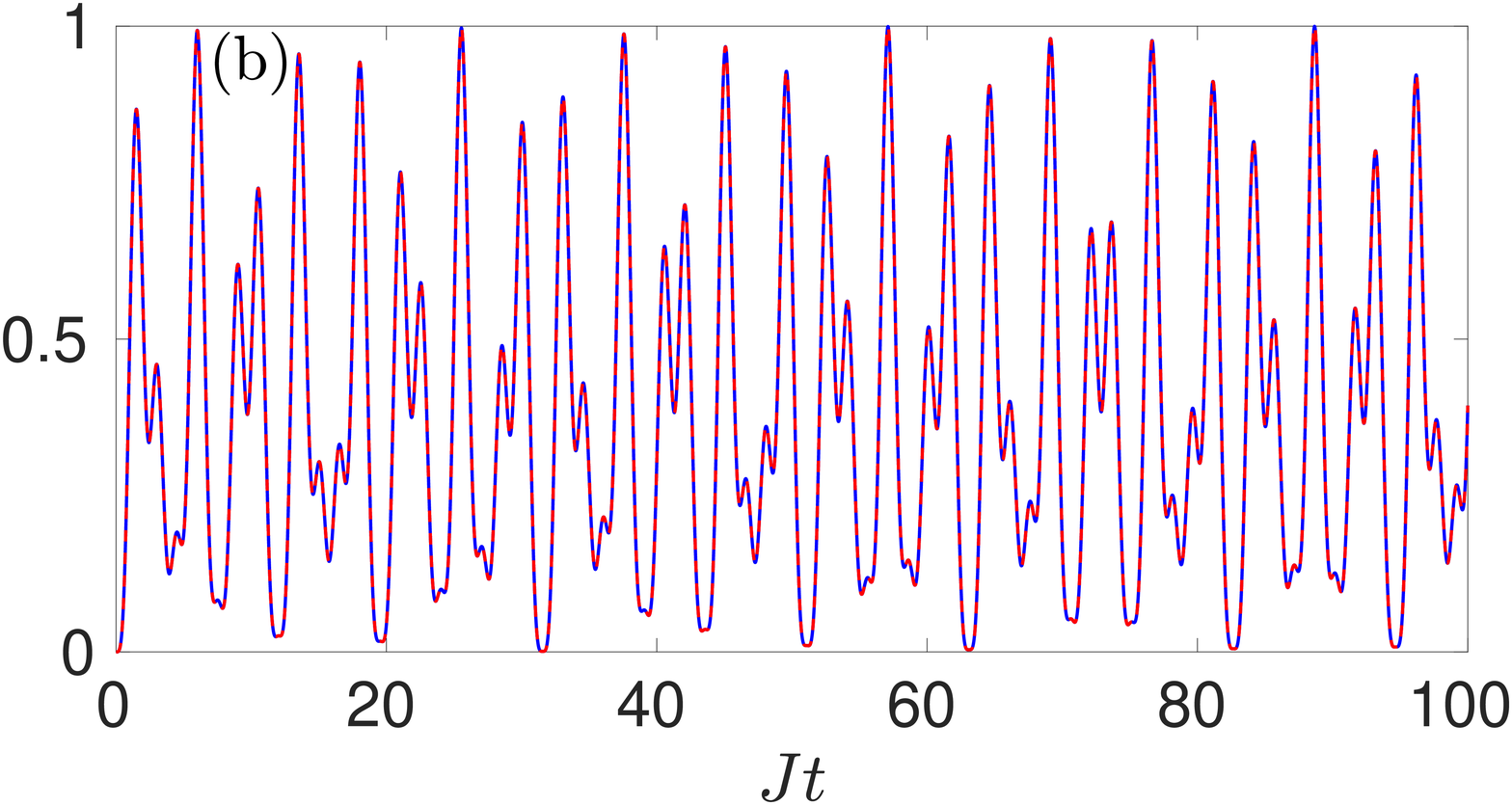}
    \end{minipage}
    \begin{minipage}{.32\textwidth}
        \centering
        \includegraphics[width=2.2in, height=1.8in]{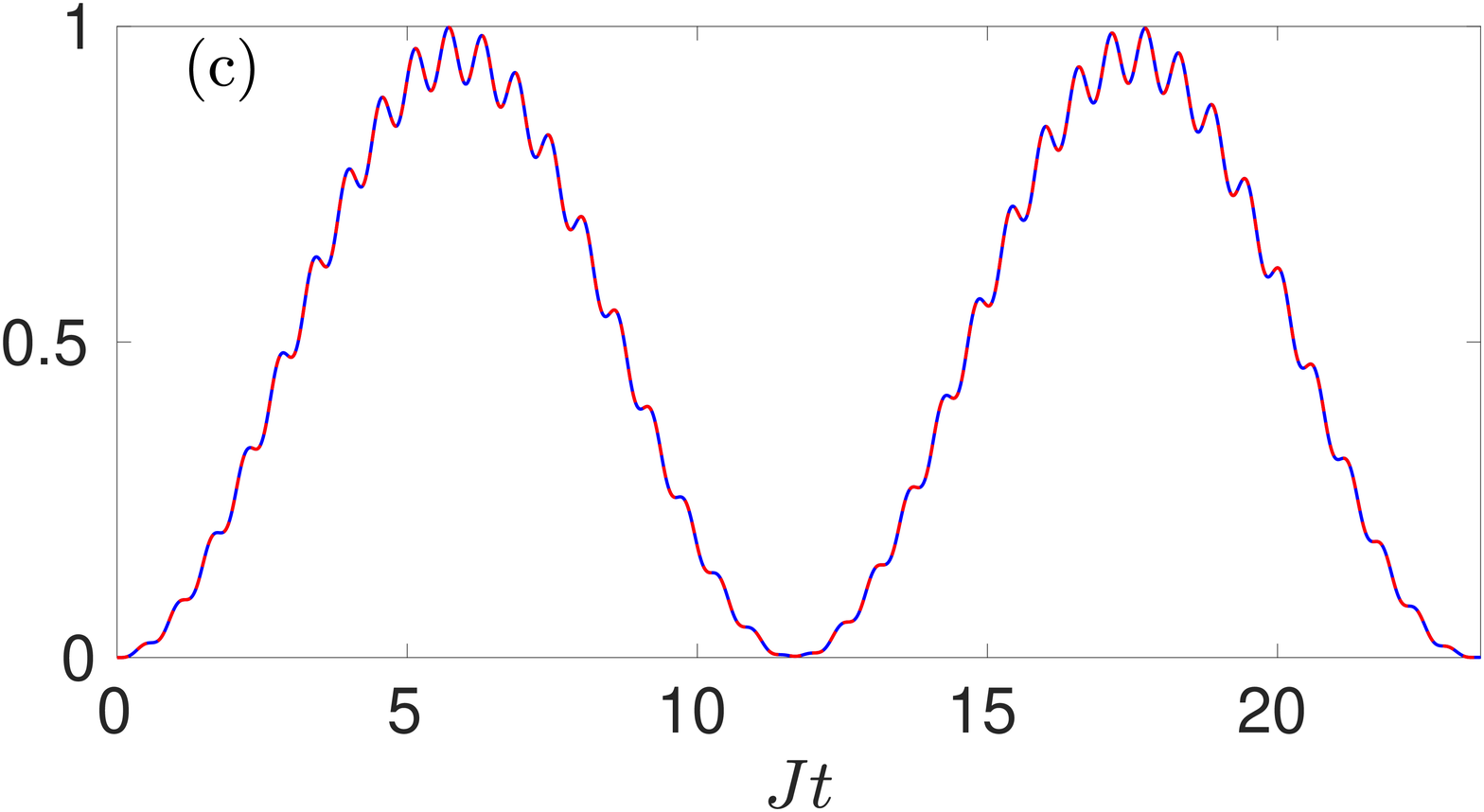}
    \end{minipage}
    \begin{minipage}{.32\textwidth}
        \centering
        \includegraphics[width=2.2in, height=1.8in]{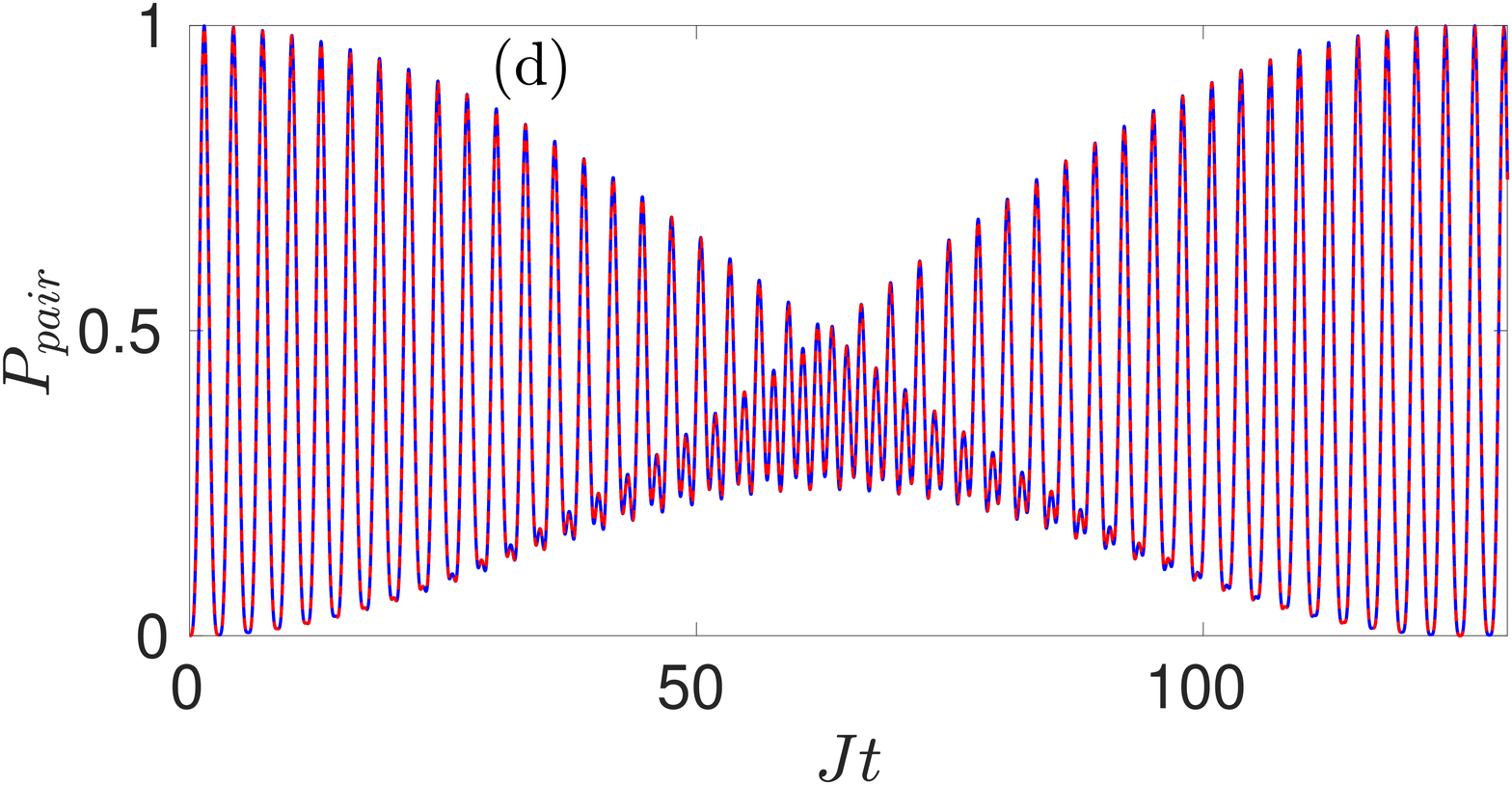}
    \end{minipage}
    \begin{minipage}{.32\textwidth}
        \centering
        \includegraphics[width=2.2in, height=1.8in]{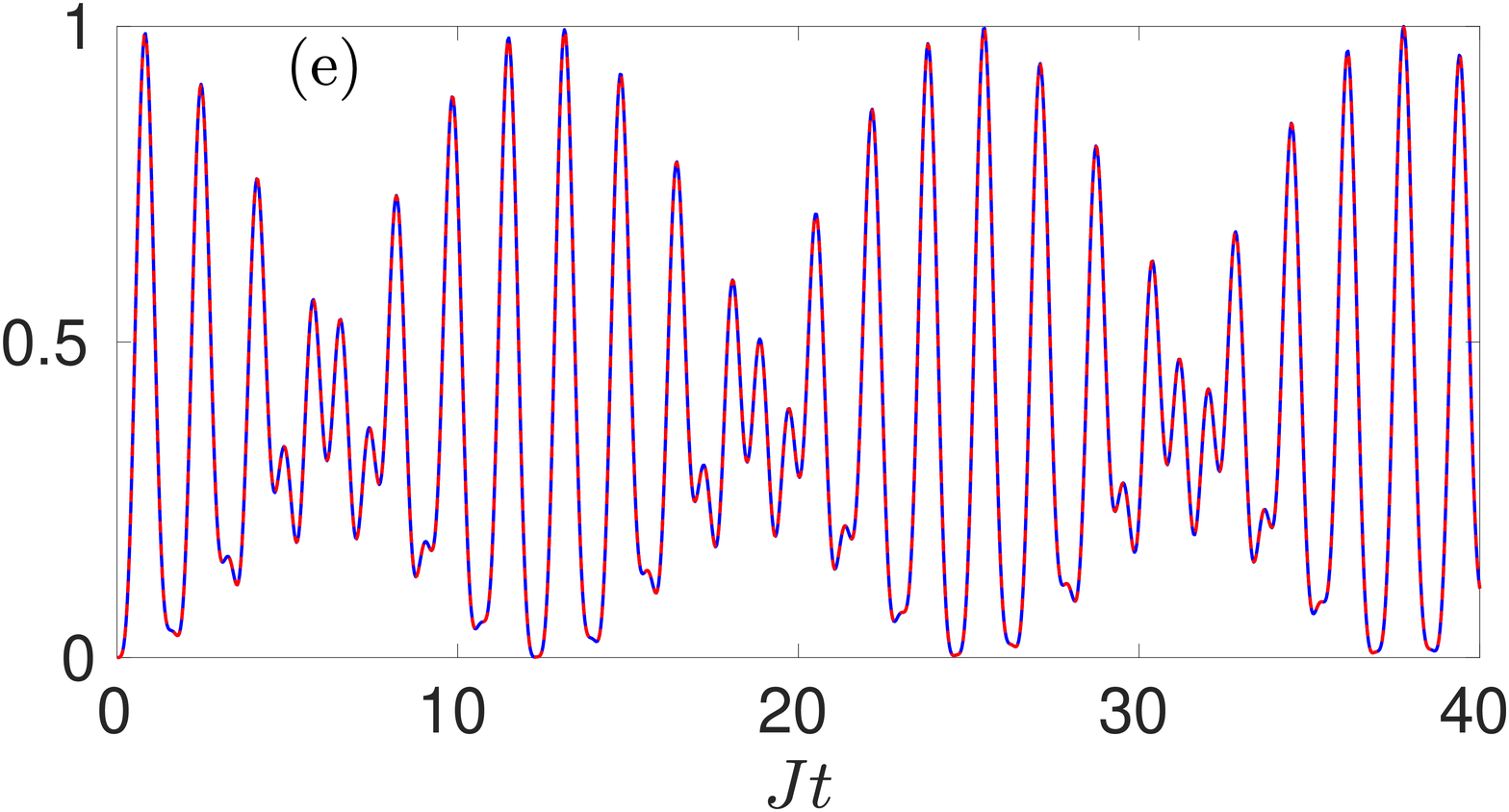}
    \end{minipage}
    \begin{minipage}{.32\textwidth}
        \centering
        \includegraphics[width=2.2in, height=1.8in]{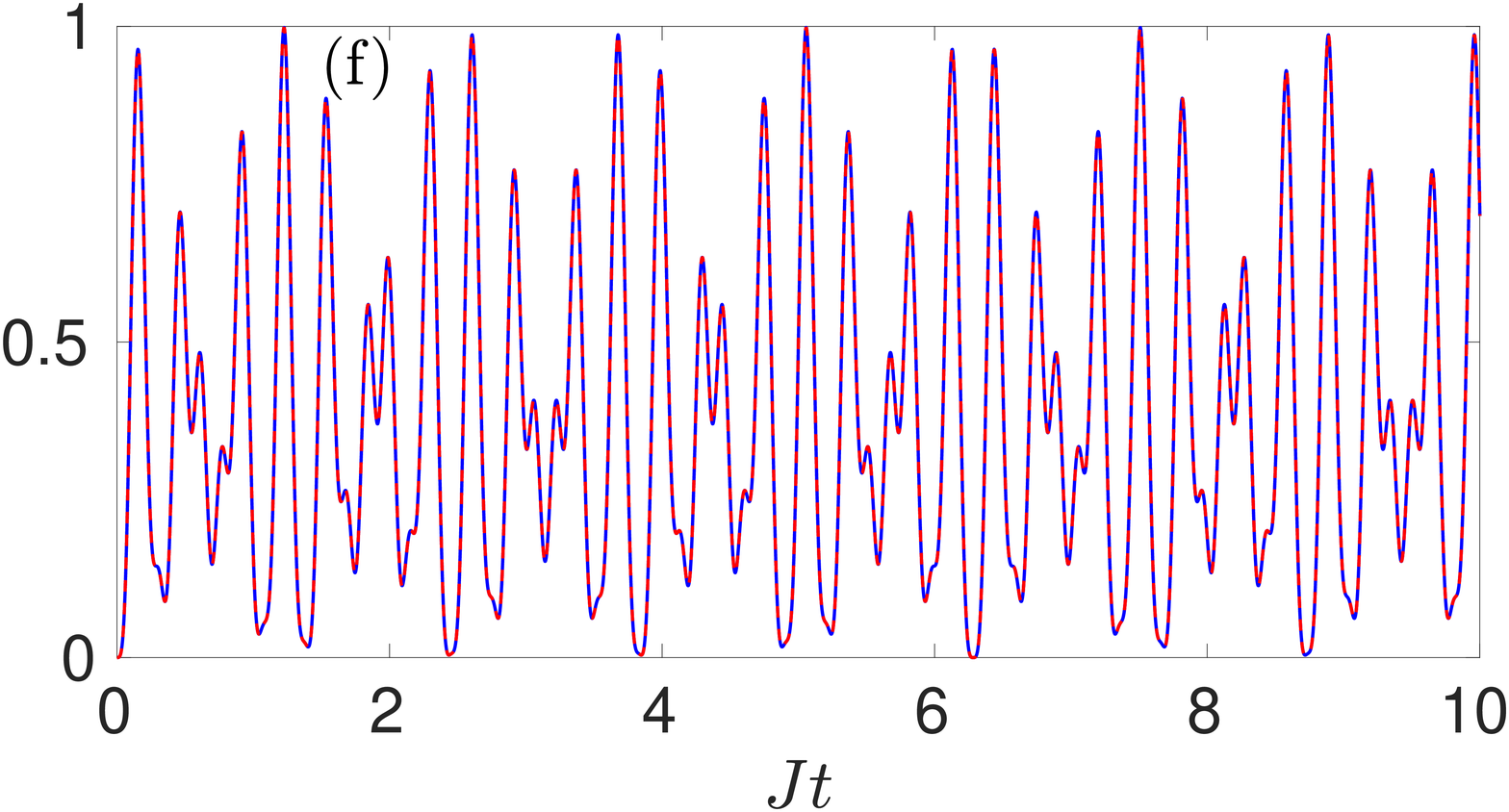}
    \end{minipage}
 \caption{\small The pair tunneling as a function of dimensionless time $Jt$ with initially both particles being in the same site. The onsite interaction parameters are $U/J=0.1$(a,d), $=1$ (b,e) and $=10$(c,f). The other interaction parameters are $U_i= U/600$, $I=-U/50$, $K=U_i$ for (a,b,c) and $U_i=U/10$, $I=-U/2$, $K=U_i$ for (d,e,f)}
 \label{Figure 5.}
\end{figure}

We define the probability of single occupancy $\rho_s$ as the probability of finding one particle in each site and the probability of double occupancy $\rho_d$ as probability of finding both particles in same site irrespective of whether both particles occupy the left or right site. They turn out to be the same for bosonic and fermionic cases for the same initial conditions. For both particles initially in one site, we obtain 
\begin{eqnarray}
\label{eq14}
 \rho_d &=& 1-\frac{8J_{-}^2}{\Omega^2}\sin^2\frac{\Omega t}{2} \\ 
 \rho_s &=& \frac{8J_{-}^2}{\Omega^2}\sin^2\frac{\Omega t}{2}
 \label{eq15}
 \end{eqnarray} 

 The above equations tell us that if $|U_{-}| >\!>J_-$, $\rho_d \simeq 1$ and $\rho_s <\!< 1$ for all times. On the other hand, for $|U_{-}| <\!< J_-$, $\rho_d$ as a function time will vary between between 1/2 and 1 and $\rho_s$ between 0 and 1/2. For the other initial condition, that is, each atom initially residing in two different wells, we have
 \begin{eqnarray}
 \rho_d &=& \frac{16J_{-}^2}{\Omega^2}\sin^2\frac{\Omega t}{2} \\ 
 \rho_s &=& 1-\frac{16J_{-}^2}{\Omega^2}\sin^2\frac{\Omega t}{2}
 \end{eqnarray} 
which show temporal behavior that is complementary to those for the former intial condition, that is, for $|U_{-}| >\!>J_-$, $\rho_d <\!<1$ and
$\rho_s \simeq 1$ while for $|U_{-}| <\!< J$, $\rho_d$ and $\rho_s$ will vary between between 0 and 1.
 
 \begin{figure}[h]
\centering
    \hspace{-.2in}
    \includegraphics[width=3.5in, height=2.2in]{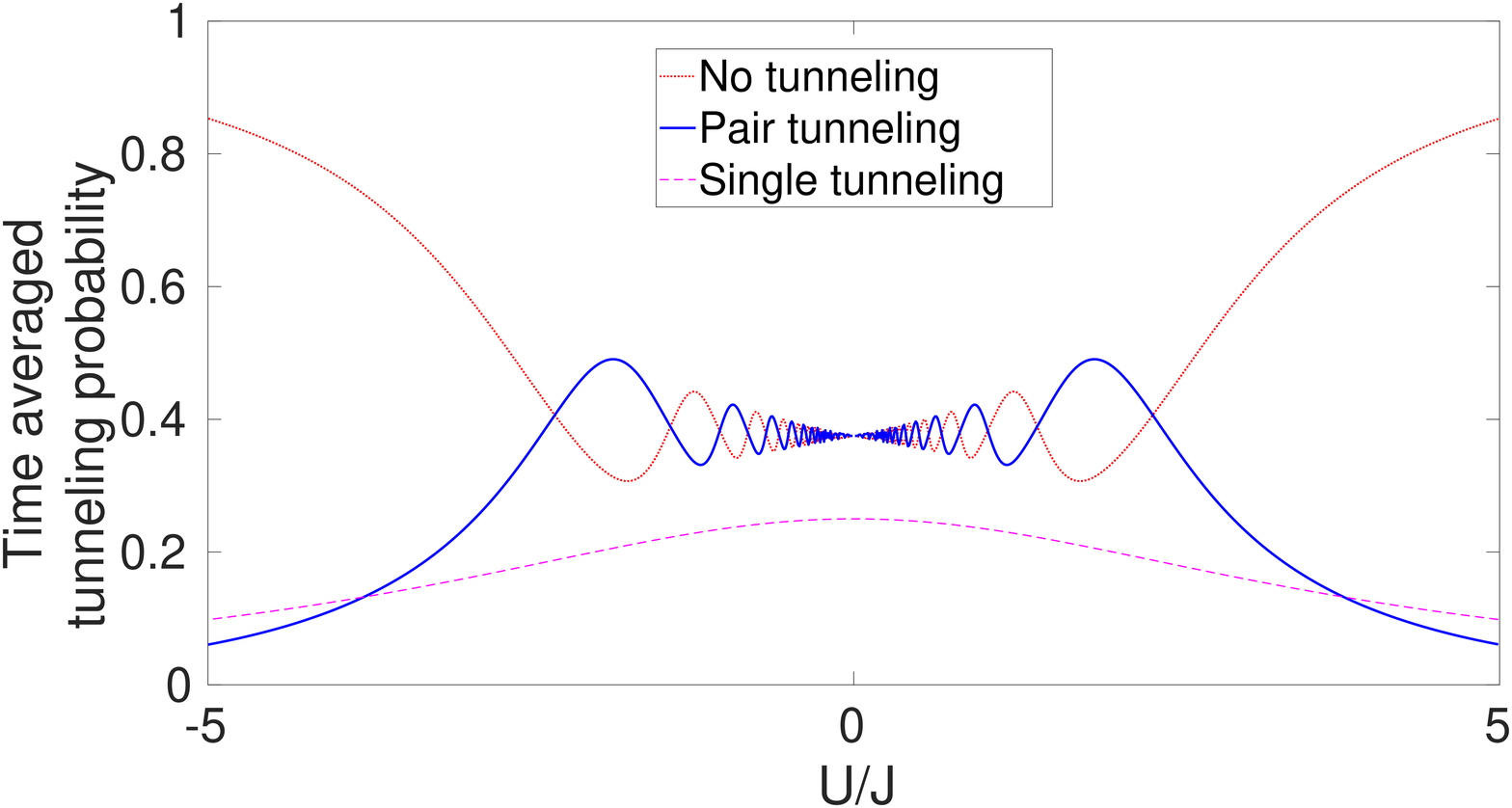}
 \caption{\small The time-averaged probabilities of pair tunneling, single particle tunneling and no tunneling are plotted as a function of $U/J$ with initially both particles in right well.}
 \label{Figure 6.}
 \end{figure}

The pair tunneling probability ($P_{pair}$) is defined as the probability of finding both atoms in the right well after initializing the system with both atoms in left well or vice versa. The single particle tunneling probability ($P_{single}$) is defined as the probability of finding one atom in each site for initially both atoms being in the right or left site; whereas, no tunneling probability is the probability of finding the initial state.  Explicitly, for initially both atoms being in the same site, we have 
\begin{eqnarray}
\label{eq16}
 P_{pair} &=& \frac{1}{4}\Bigg\{\left(\frac{3}{2}+\frac{U_{-}^2}{2\Omega^2}\right) - \left(1+\frac{U_{-}}{\Omega}\right)\cos\frac{(\Omega-U_{-}) t}{2}\nonumber \\
 &-& \left(1-\frac{U_{-}}{\Omega}\right)\cos\frac{(\Omega+U_{-}) t}{2}\nonumber\\ &+& \frac{1}{2}\left(1-\frac{U_{-}^2}{\Omega^2}\right)\cos\Omega t \Bigg\} \\
 P_{single} &=& \frac{8{J_{-}}^2}{\Omega^2}\sin^2\frac{\Omega t}{2}
 \label{eq17}
\end{eqnarray}
Again, these tunneling probabilities are  same for both bosonic and fermionic cases for the same input parameters.

 When $U_i \ne 0$ and  the system is prepared with both particles initially in left (or right) well, the time evolution of pair tunneling probability as shown in Fig.\ref{Figure 5.} has more than one frequencies of oscillations as the analytical result of Eq.(\ref{eq16}) reveals. For $U\neq 0$ and $ U_i \ne 0 $, $P_{pair}$ has three frequencies $(\Omega-U_{-})/4\pi, (\Omega+U_{-})/4\pi$ and $\Omega/2\pi$. From  Eqs.(\ref{eq16},\ref{eq17}), we find that for $U=0=U_i$ as in Fig.\ref{Figure 5.}a, there are two frequencies of $P_{ pair}$ one is $J/\pi$ and other is $2J/\pi$. On the other hand for $U>\!>J$, there is single frequency dominance with very little modulation and the frequency being $U_-/2\pi$ (Fig.\ref{Figure 5.}c). In the intermediate interaction where $U=J$, there remains all three frequencies as shown in Fig.\ref{Figure 5.}b. Now we consider the other interaction parameters ($U_i, I, K$) to be relatively large which makes them comparable to the on-site interaction $U$. Then we plot the pair tunneling (Fig.\ref{Figure 5.}d,e,f) and see the effect of these interaction terms. The change in modulation frequency is easily noticed for lower values of interaction. For strong interacting case, instead of single-frequency dominance, we find modulation. This represents that the other interaction terms largely modifies the tunneling properties if the magnitude of those interactions are comparable to $U$. We take the time average of the  tunneling probability over the  period $T_{max}$.  We have plotted the time-averaged probabilities of single-particle  and pair tunneling as a function of $U$ in Fig.\ref{Figure 6.} for the system initially prepared with two atoms in single site. This shows the fact that when the interaction is 
sufficiently large, the system has tendency to stay in the same state that it was initially prepared.
 
We next discuss entanglement and quantum fluctuations for  the two systems.

\section{Quantum entanglement and quantum fluctuation}\label{5}
 
\subsection{Entanglement}\label{5.1}

The states $|s\rangle$ and $|t\rangle$ given by Eq.(\ref{eq:s}) and Eq.(\ref{eq:t}) are two of four well-known Bell basis in spin variables. Similarly, the states $\mid\pm\rangle$, $\mid0\rangle$ and $\mid1\rangle$ may be considered as four Bell basis of spatial variables. The entanglement properties of a bipartite system consisting of two subsystems $\alpha$ and $\beta$ can be studied by calculating the von Neumann entropy of the reduced density matrix $\rho_{\alpha} = {\rm Tr}_{\beta} [ \rho_{\alpha\beta}]$ or $\rho_{\beta} = {\rm Tr}_{\alpha} [ \rho_{\alpha\beta}]$ where $\rho_{\alpha\beta}$ is the joint density matrix of the bipartite system and ${\rm Tr}_{\alpha(\beta)}$ implies trace over the subsystem $\alpha(\beta)$. In the present context, one can partition the system into two possible ways: One in terms of the two spatial (localized) modes corresponding to the left (L) or right (R) sites ($\alpha, \beta \equiv L, R$) of the double-well, another in terms of the complete wave functions of two-particle system described in terms of their individual spin and position degrees-of-freedom (DOF) or in second quantization formalism. In the latter case, one needs to calculate the reduced one-particle density matrix $\rho_1={\rm Tr}_2[\rho_{s}]$, where $\rho_{s}=\mid\psi_{s}\rangle\langle\psi_{s}\mid$ is the two-particle density matrix for fermions ($s \equiv F$) or bosons ($s \equiv B$). Here ${\rm Tr}_2$ implies trace over all the basis states of particle 2.

  \begin{figure}[h]
        \includegraphics[width=3.6in, height=2.4in]{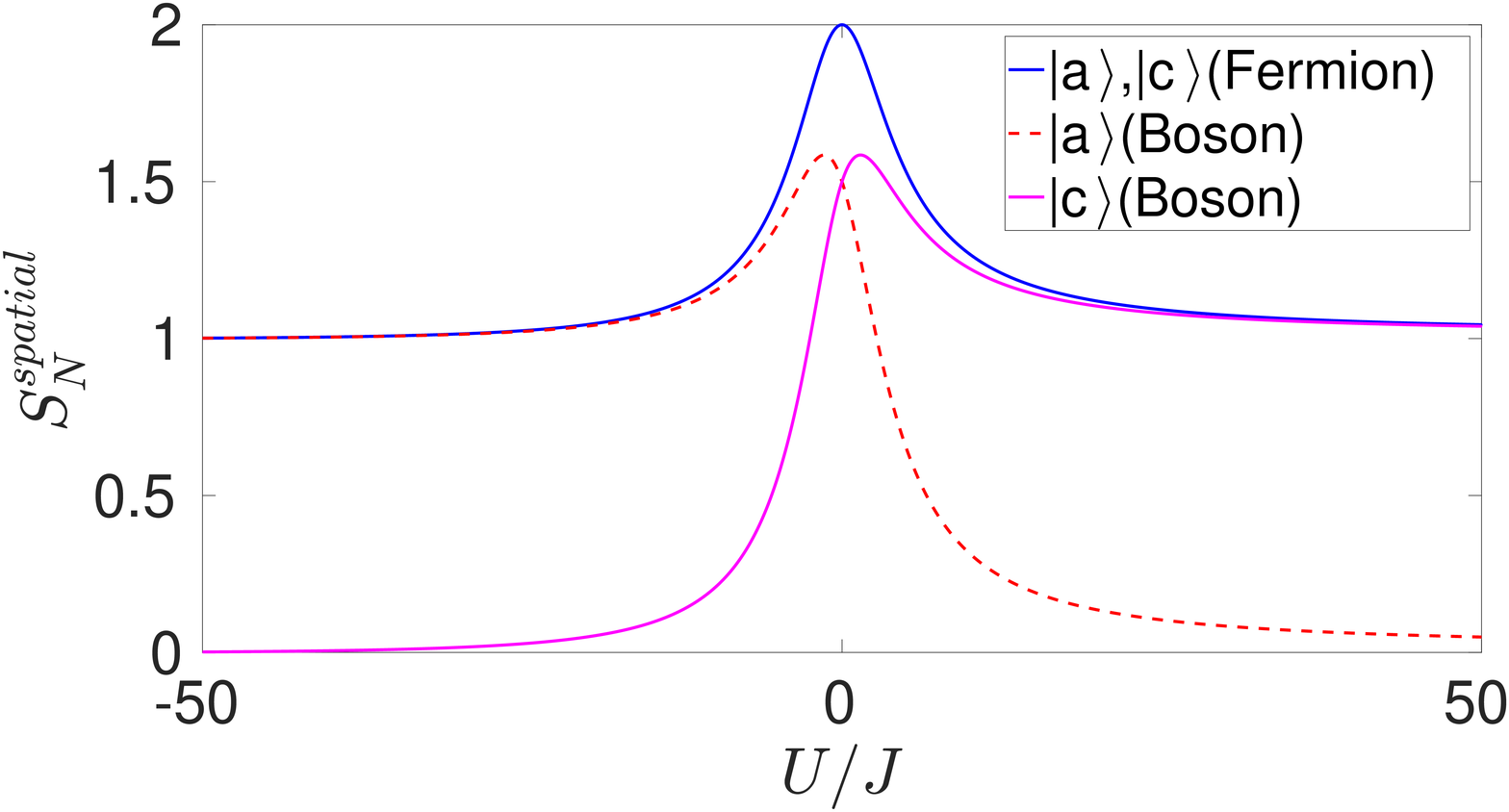}
 \caption{\small Variation of spatial mode entanglement of the eigen states with $U/J$.}
 \label{Figure 7.}
\end{figure}

  \begin{figure}[h]
    \begin{minipage}{0.32\textwidth}
        \includegraphics[width=2.2in, height=1.8in]{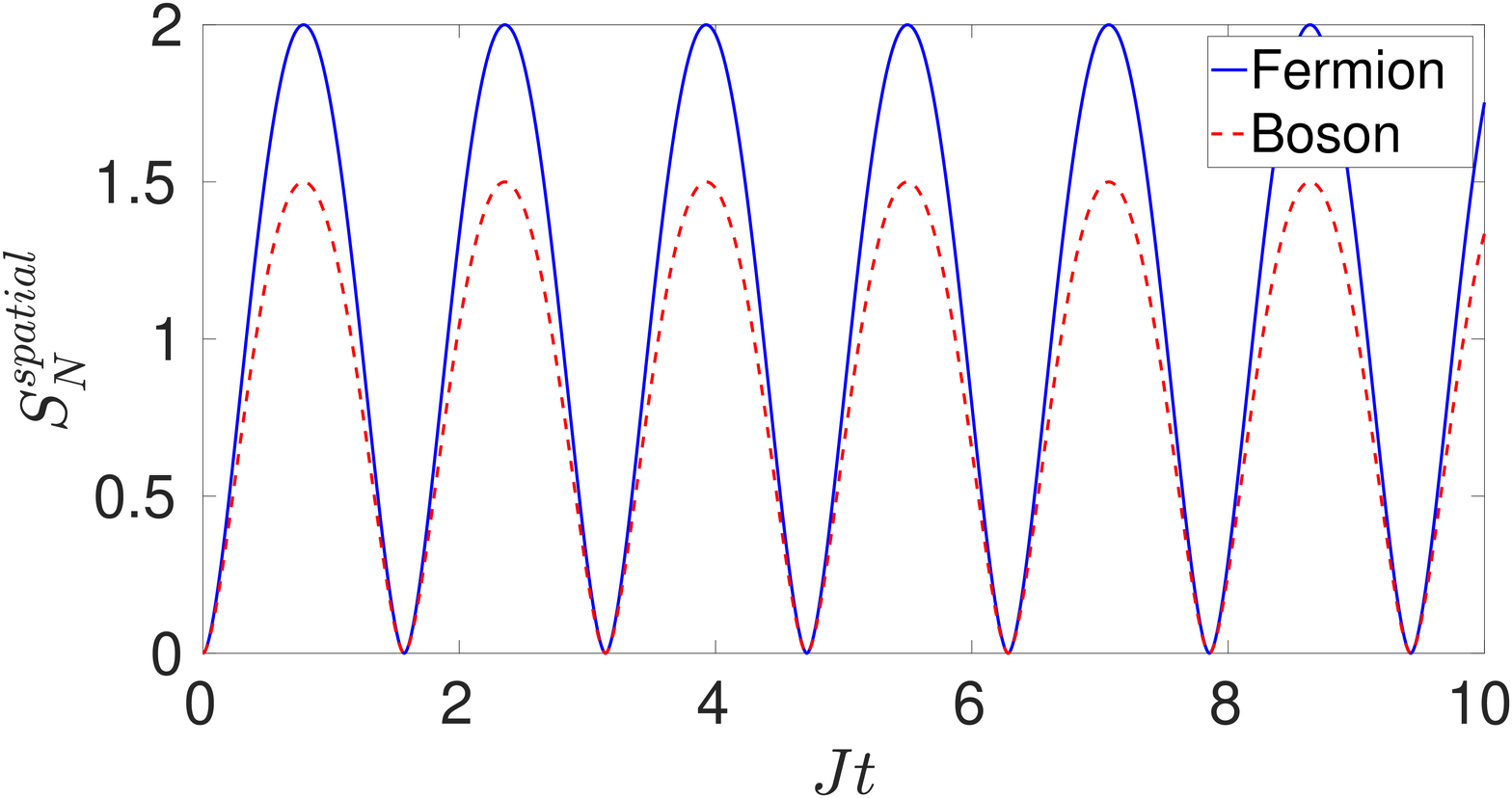}
    \end{minipage}
    \begin{minipage}{0.32\textwidth}
        \includegraphics[width=2.2in, height=1.8in]{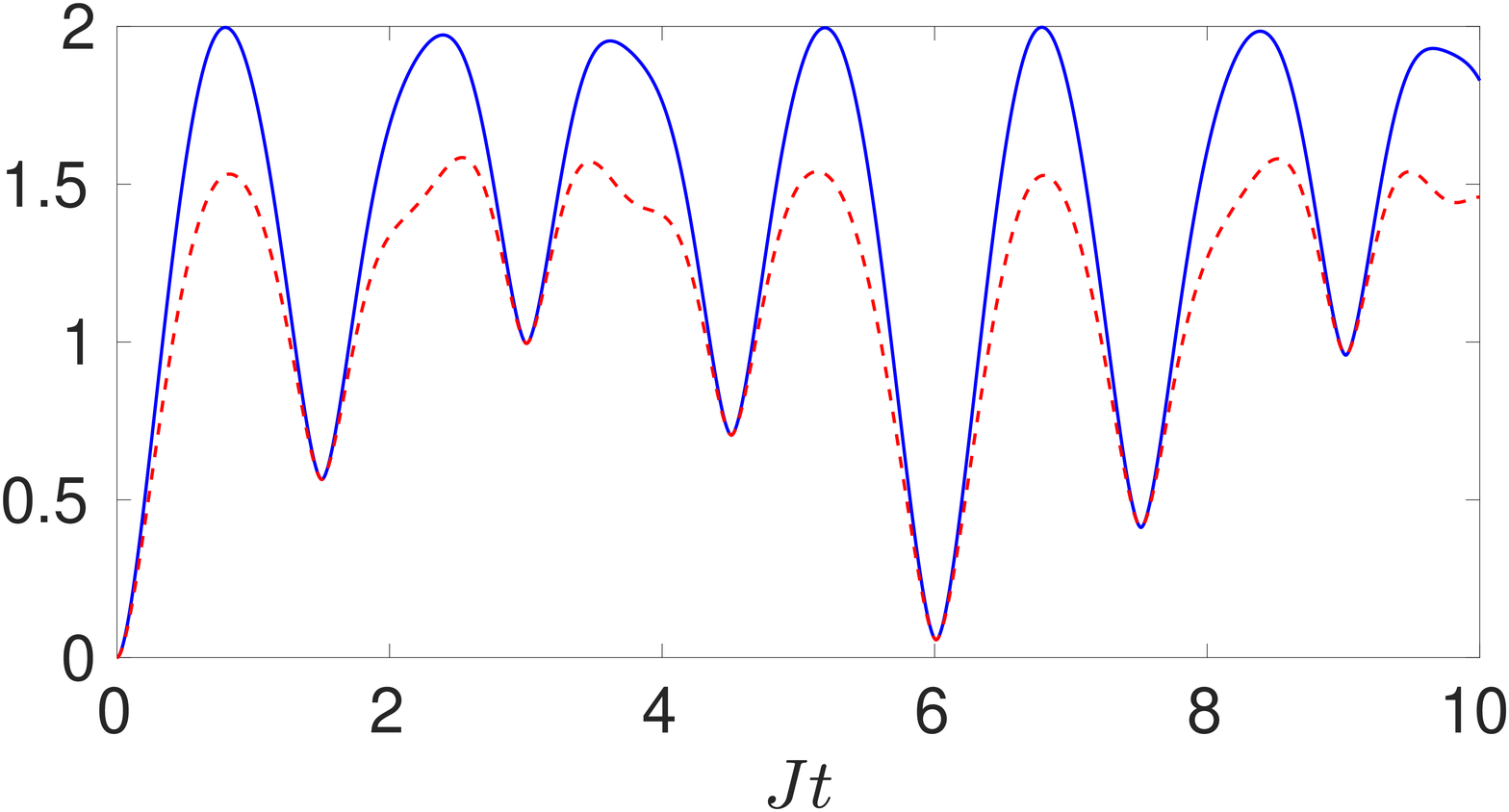}
    \end{minipage}
        \begin{minipage}{0.32\textwidth}
        \includegraphics[width=2.2in, height=1.8in]{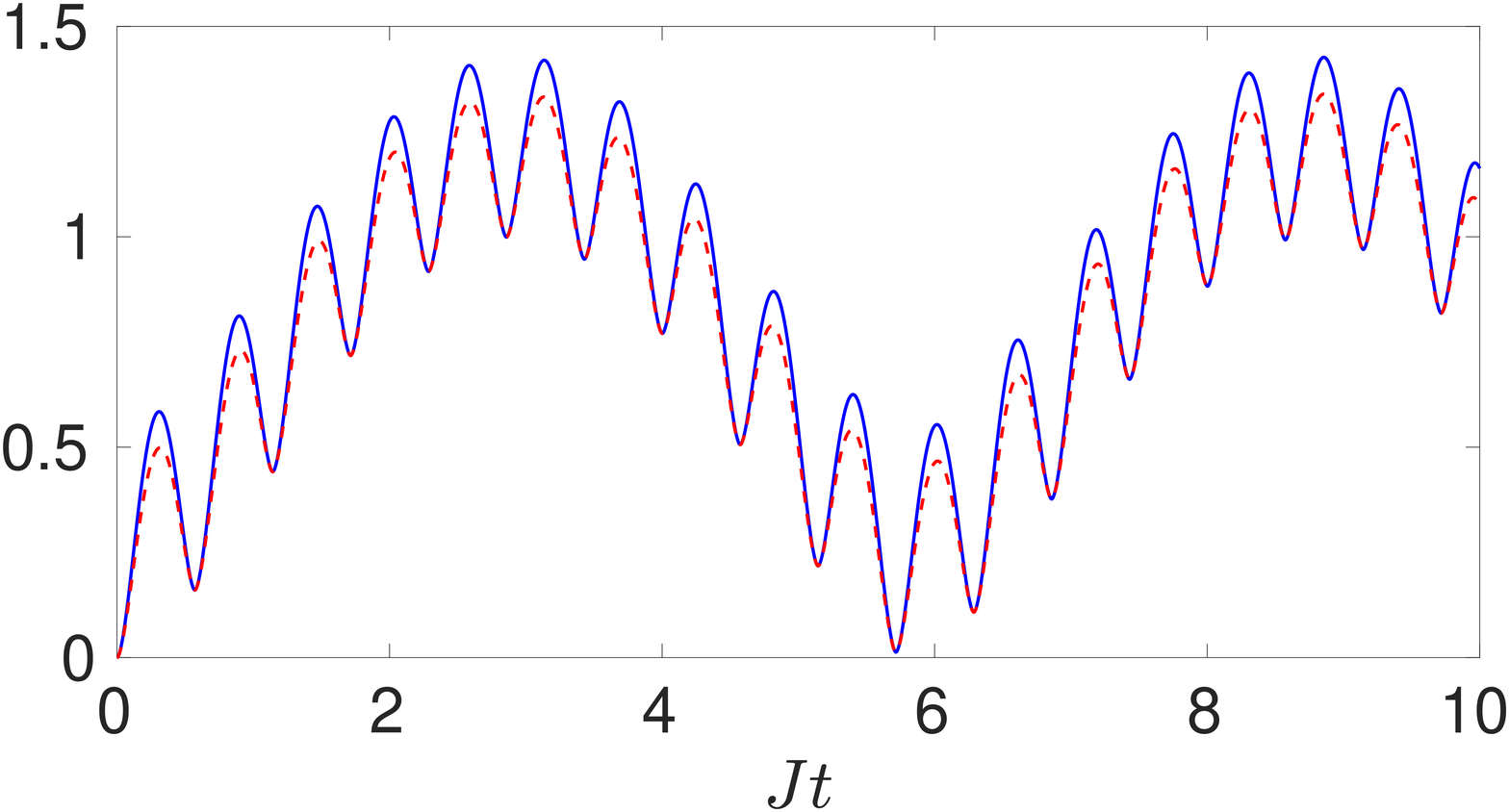}
        \end{minipage}
    \begin{minipage}{0.32\textwidth}
        \includegraphics[width=2.2in, height=1.8in]{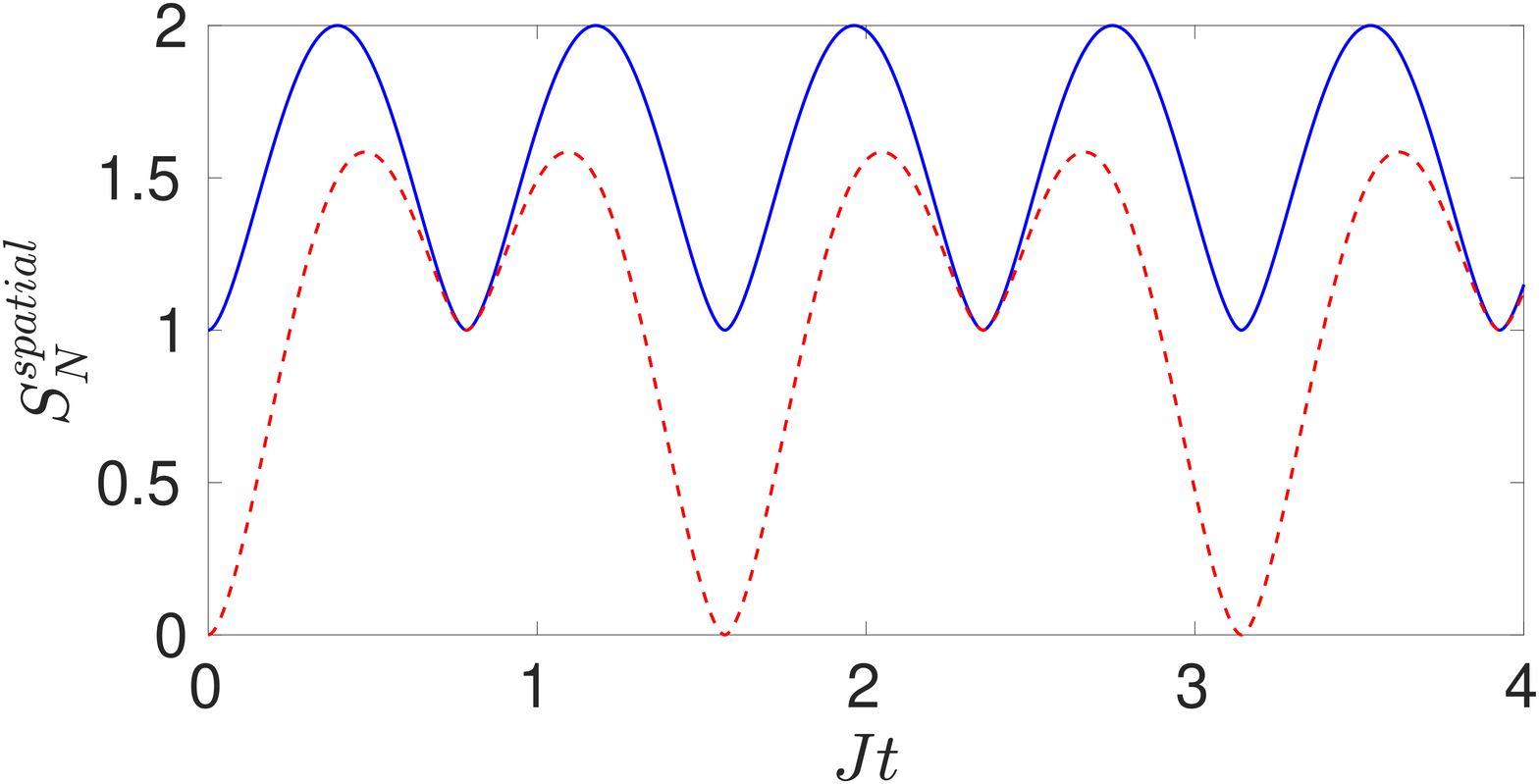}
    \end{minipage}
    \begin{minipage}{0.32\textwidth}
        \includegraphics[width=2.2in, height=1.8in]{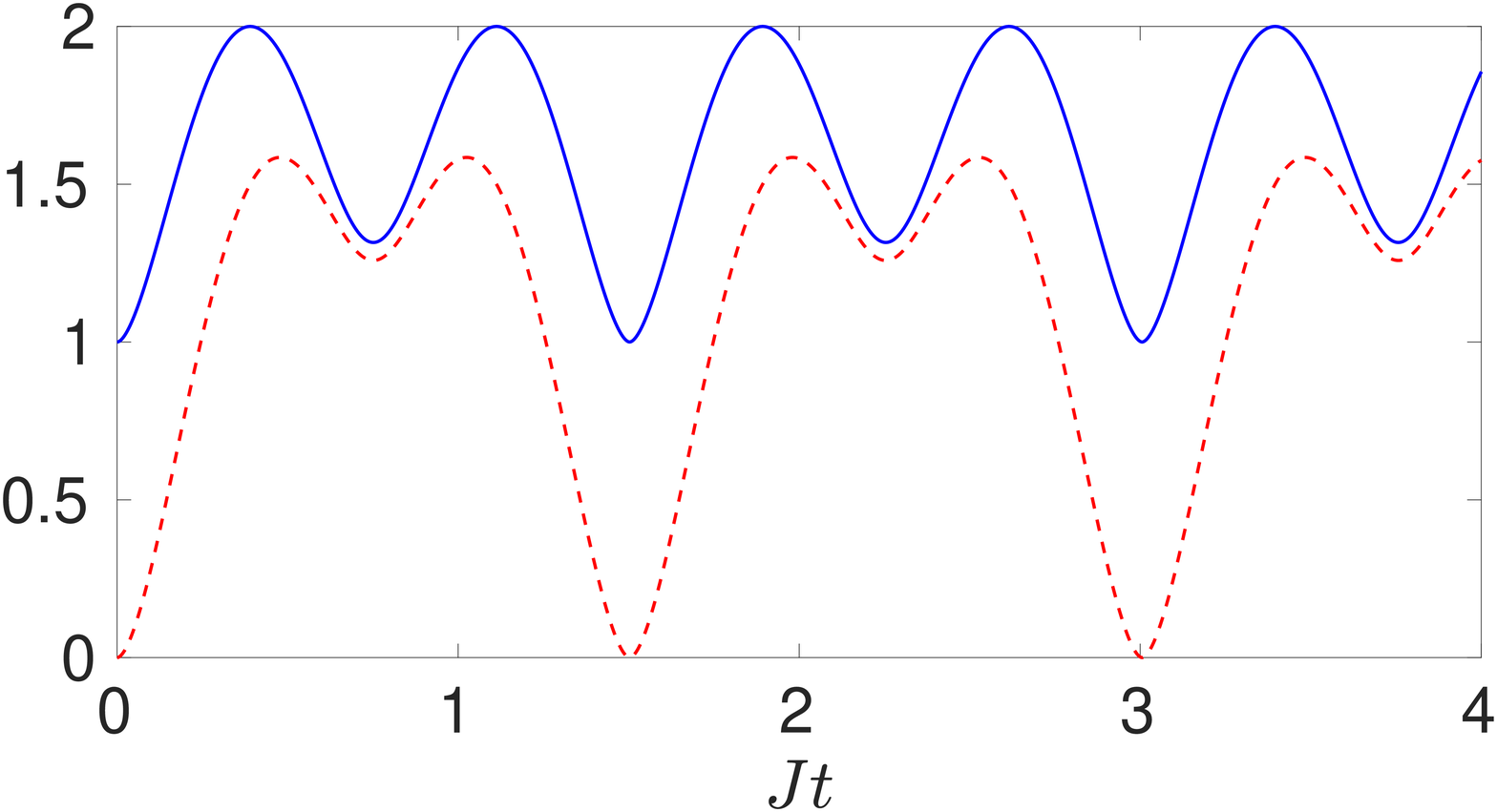}
    \end{minipage}
        \begin{minipage}{0.32\textwidth}
        \includegraphics[width=2.2in, height=1.8in]{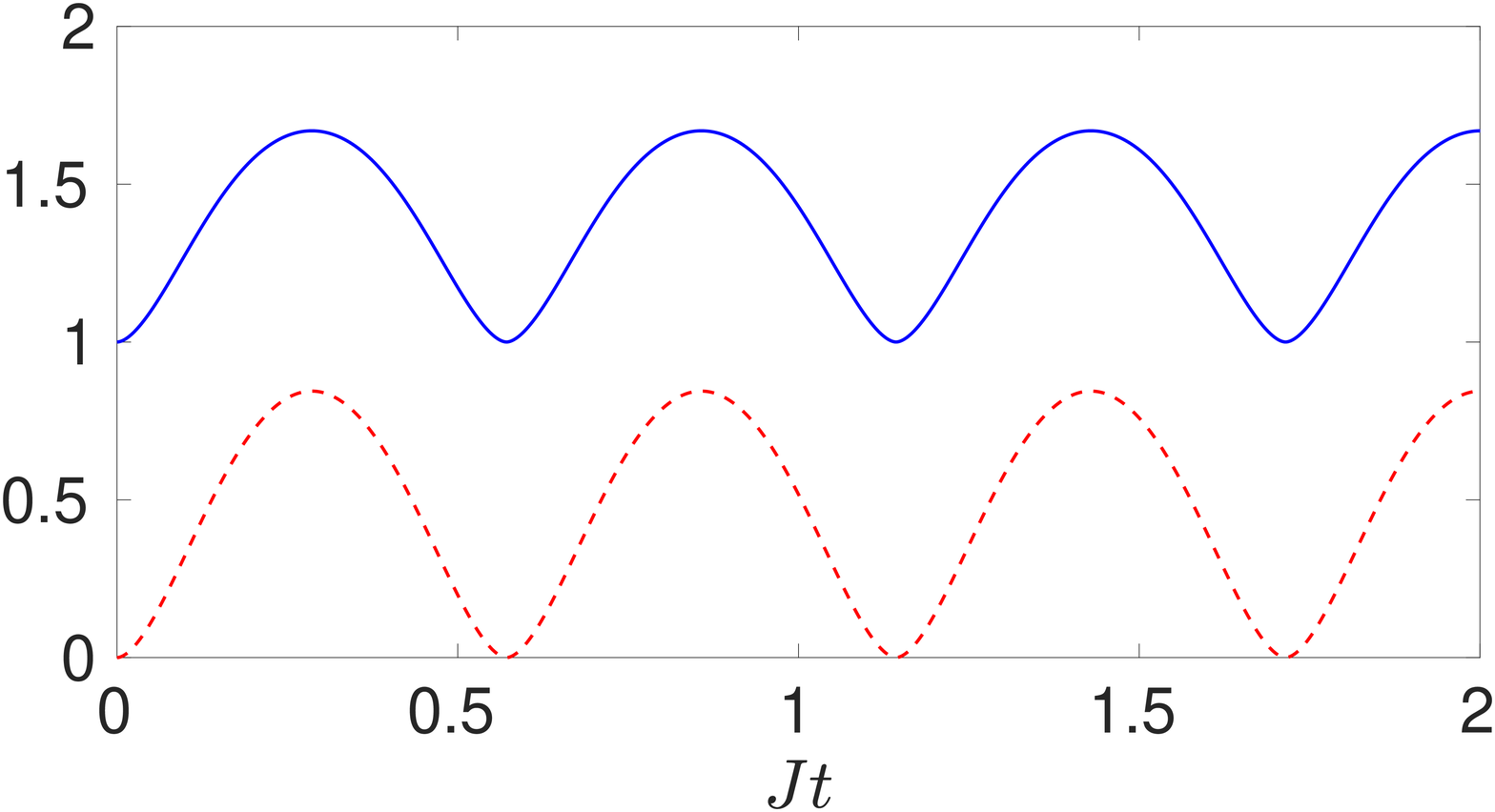}
        \end{minipage}
 \caption{\small The spatial-mode von Neumann entropy $S_N^{{\rm spatial}}$ as a function of $J t$. Form left to right: $U/J=0, 1, 10$. For the upper panel, two particles are initialized in the same well whereas for the lower panel they are initially in two different sites. Other interaction parameters are: $U_i=U/600$, $I=-U/50$ and $K=U_i$.}
 \label{Figure 8.}
\end{figure}

We calculate the von Neumann entropy $S_N^{{\rm spatial}}=-{\rm Tr}[\rho_{\alpha}{\rm log}_2(\rho_{\alpha})] = {\rm Tr}[\rho_{\beta}{\rm log}_2(\rho_{\beta})]$ of the reduced single-mode density matrix. The two spatial modes may be  entangled 
if $S_N^{{\rm spatial}}$ is nonzero. On the other hand, whether the two atoms are entangled in other DOF or whether their joint wave function $\mid\psi_{s}\rangle$ is inseparable or not can be ascertained by calculating the   
the von Neumann entropy $S_N^{(1)}=-{\rm Tr}[\rho_1{\rm log}_2(\rho_1)]$ of the 
single-particle density matrix $\rho_1$. Fig.\ref{Figure 7.} shows the variation of $S_N^{{\rm spatial}}$ of the two eigenstates of $\mid a \rangle $ and $\mid c \rangle$ of fermionic and bosonic systems as given by Eqs.(\ref{eq:eigena},\ref{eq:eigenc})
as a function of the dimensionless on-site interaction $U/J$. For the remaining two other eigenstates $\mid b \rangle$ and $\mid d \rangle$ of the fermionic system and $\mid b \rangle$ of the bosonic system,  
$S_N^{{\rm spatial}}$ is independent of $U$ and is equal to unity for both the systems. 

\begin{figure}[h]
    \begin{minipage}{0.32\textwidth}
        \includegraphics[width=2.2in, height=1.8in]{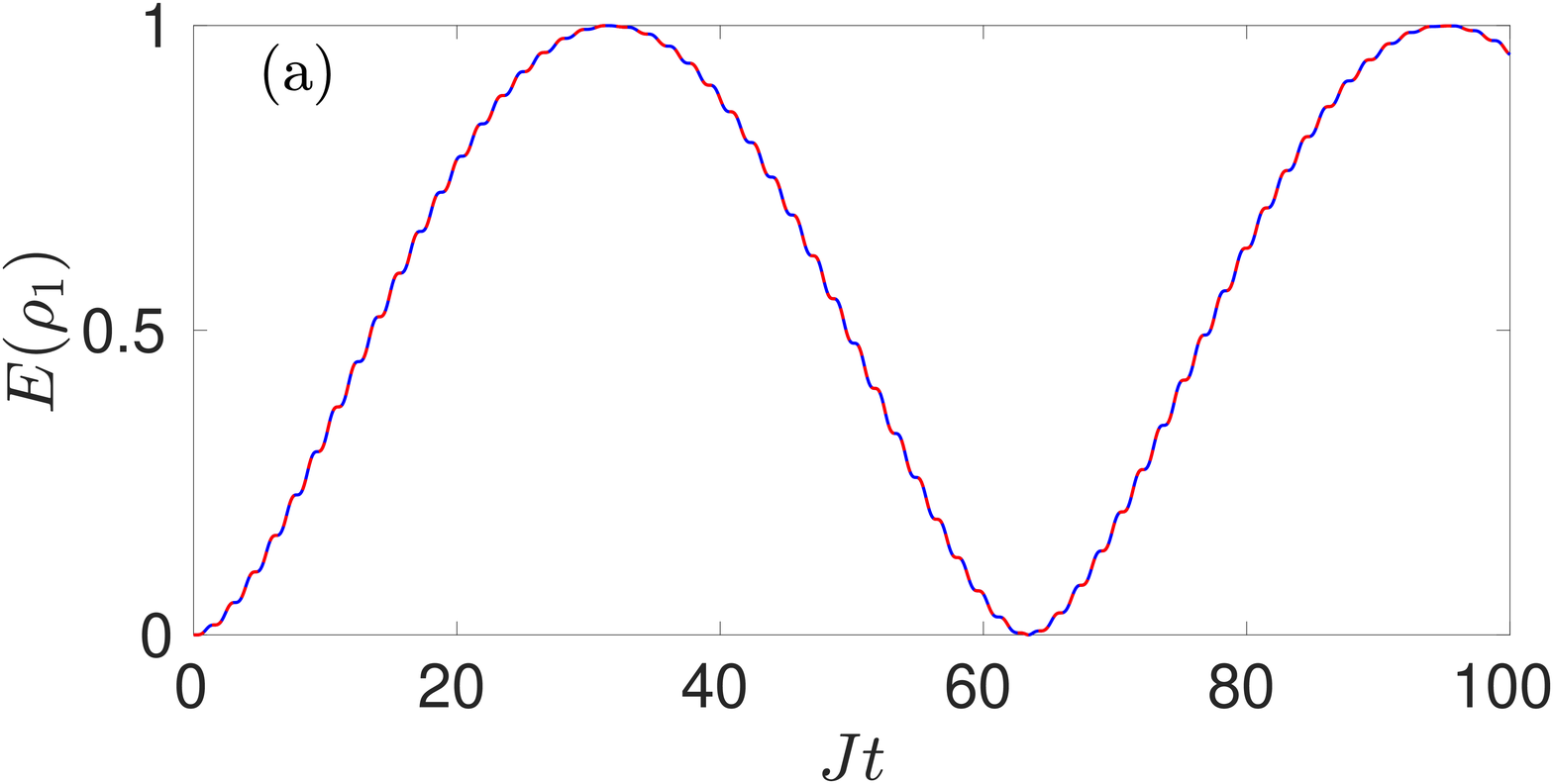}
    \end{minipage}
    \begin{minipage}{0.32\textwidth}
        \includegraphics[width=2.2in, height=1.8in]{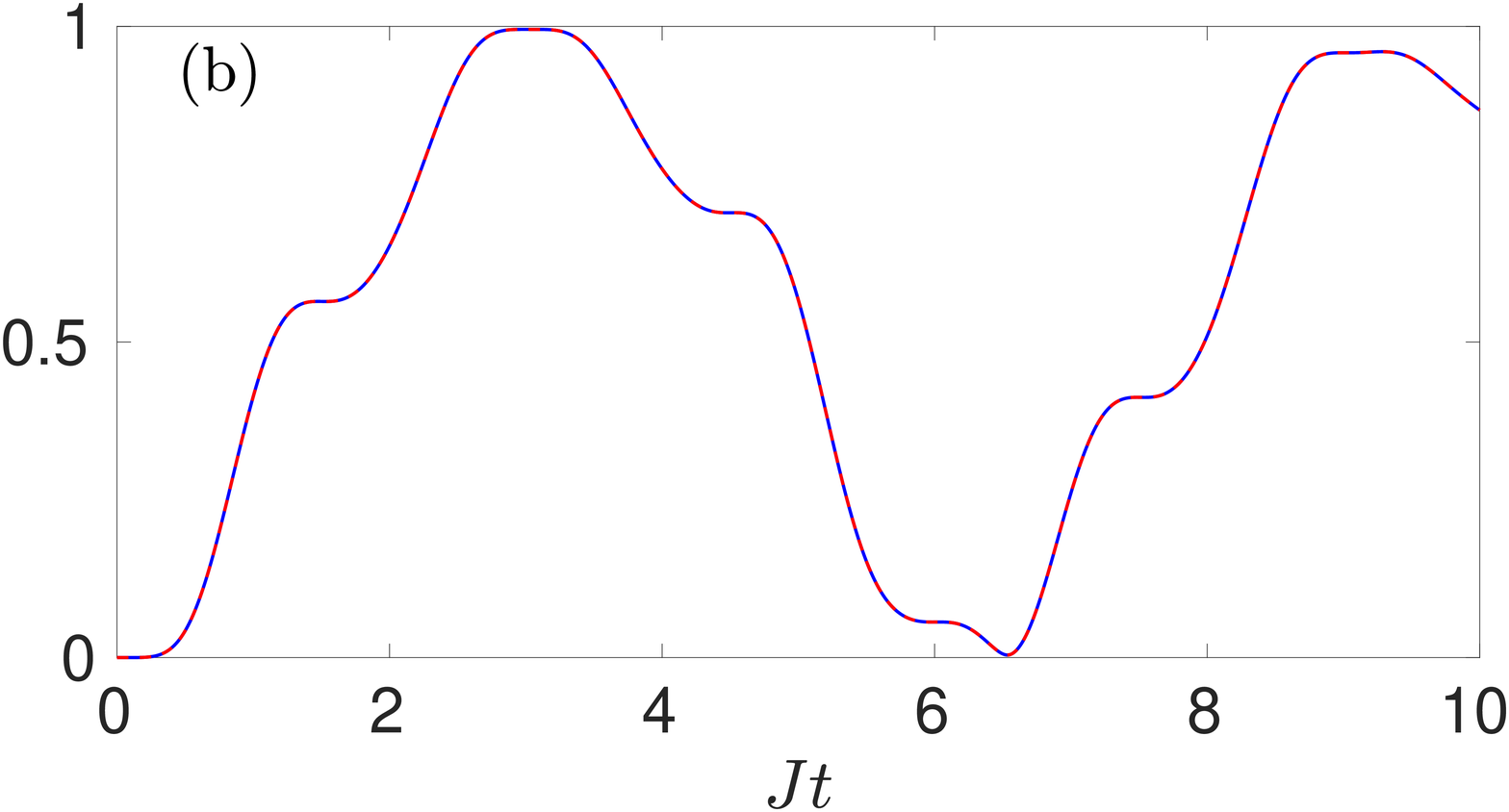}
    \end{minipage}
        \begin{minipage}{0.32\textwidth}
        \includegraphics[width=2.2in, height=1.8in]{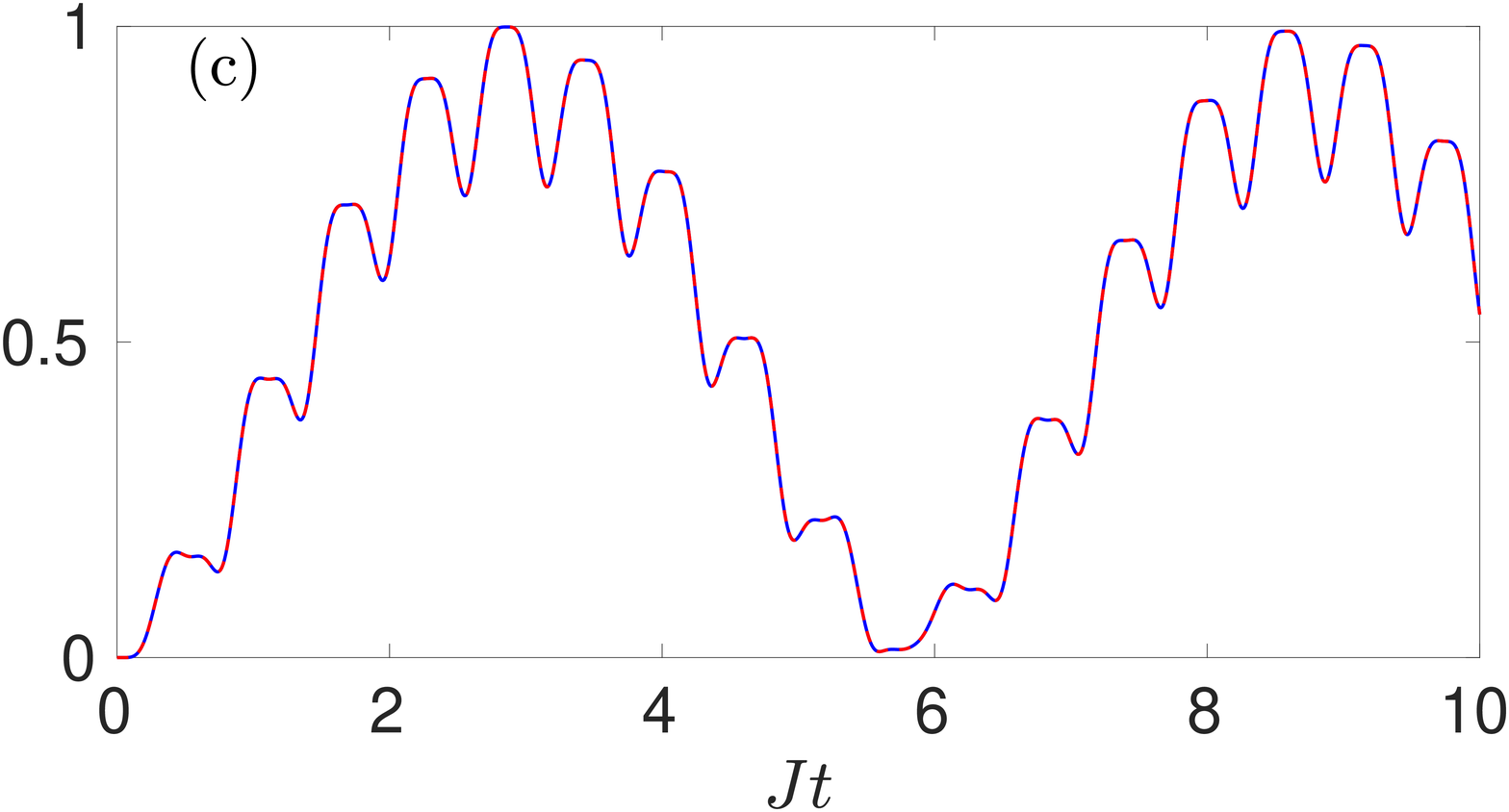}
        \end{minipage}
    \begin{minipage}{0.32\textwidth}
        \includegraphics[width=2.2in, height=1.8in]{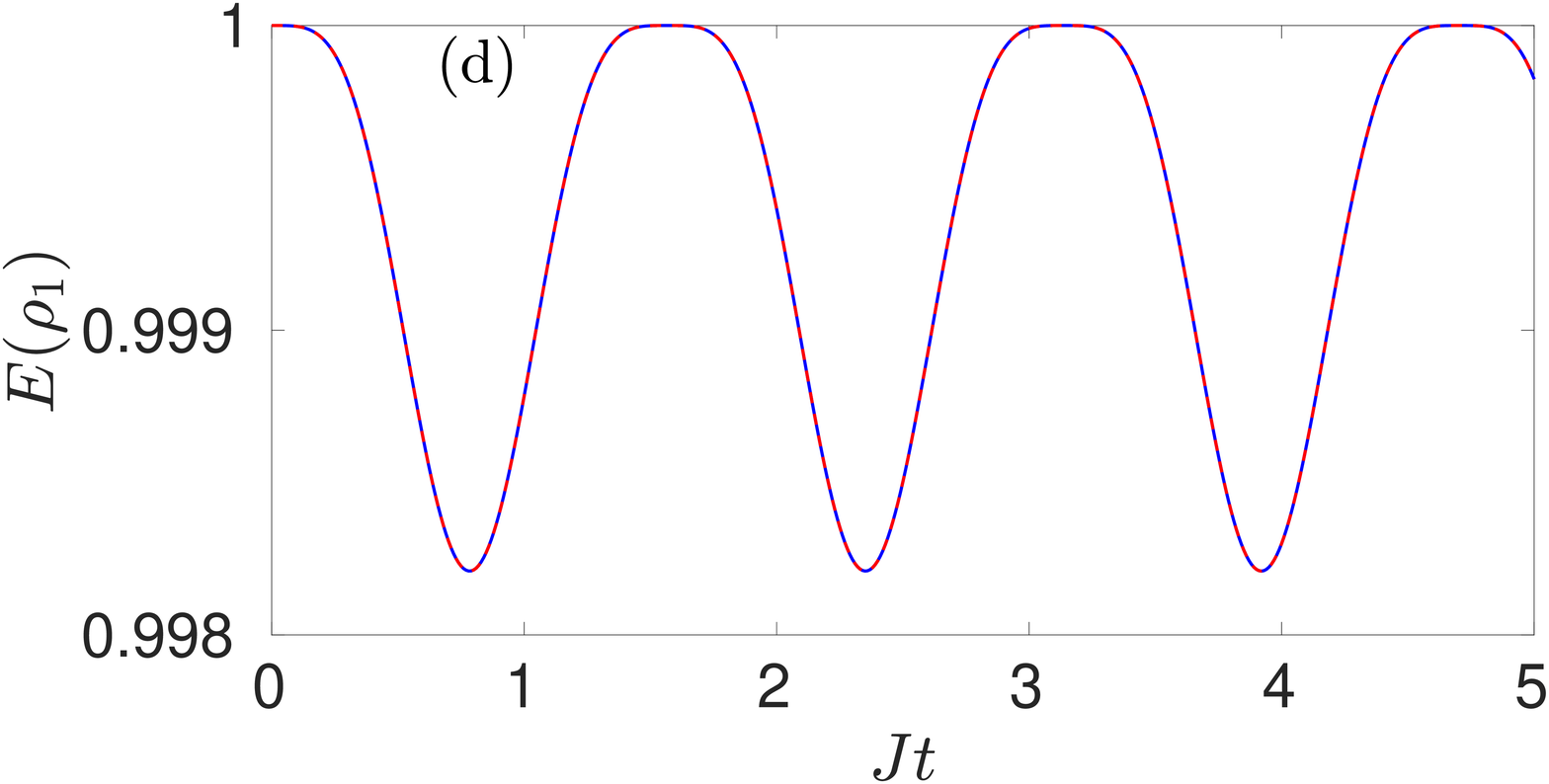}
    \end{minipage}
    \begin{minipage}{0.32\textwidth}
        \includegraphics[width=2.2in, height=1.8in]{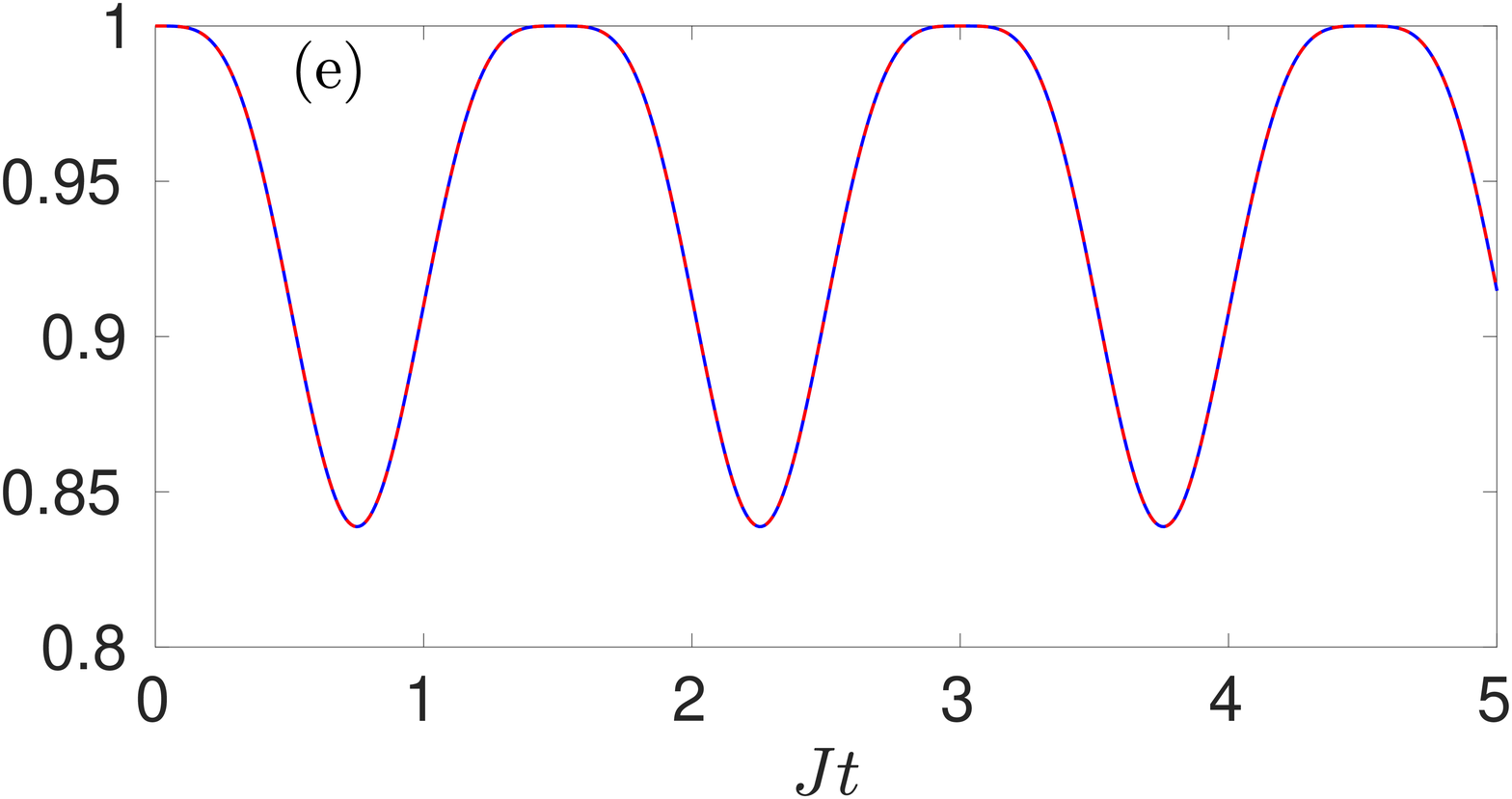}
    \end{minipage}
        \begin{minipage}{0.32\textwidth}
        \includegraphics[width=2.2in, height=1.8in]{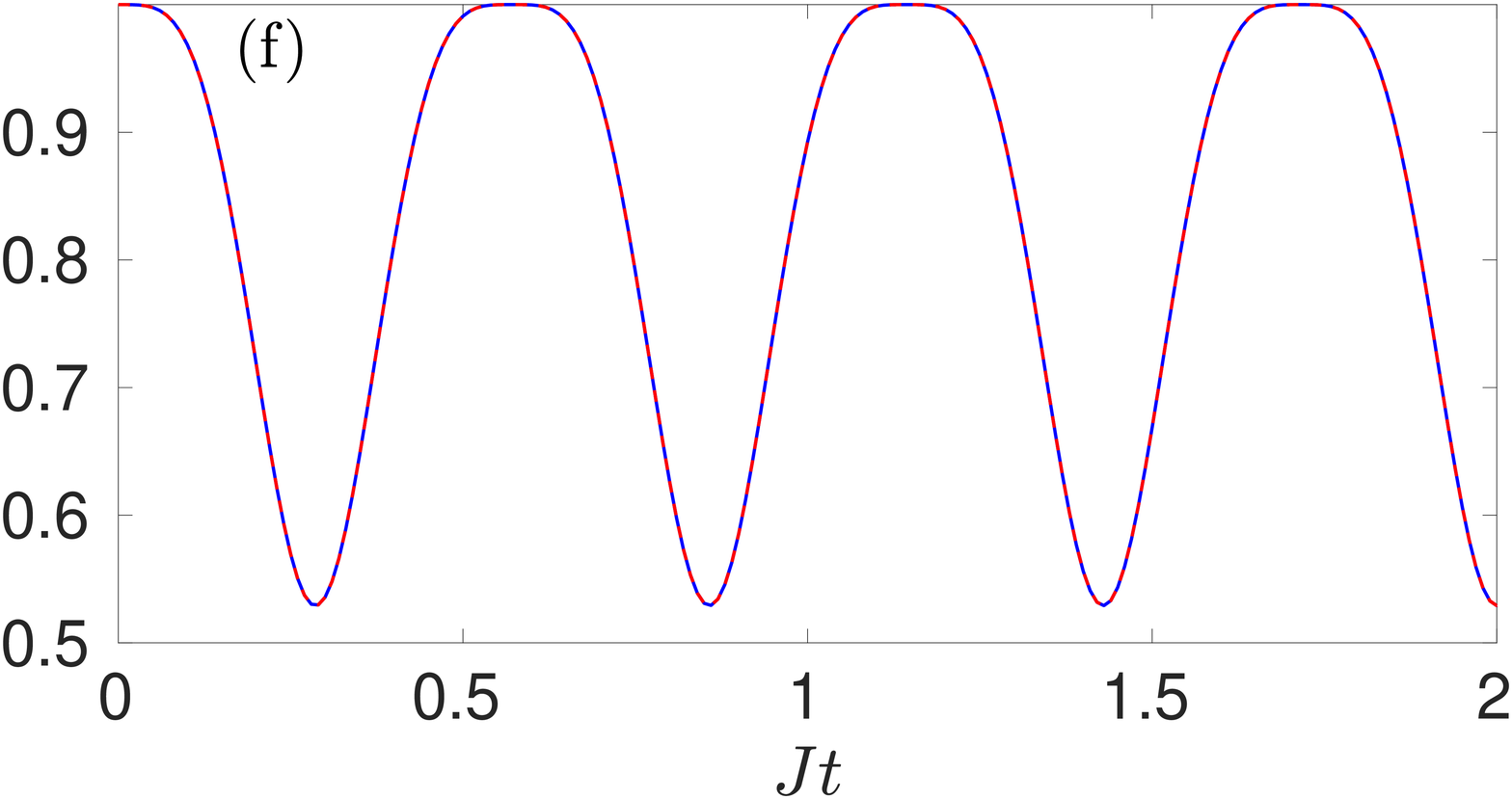}
        \end{minipage}
 \caption{\small Variation of entanglement measure $E(\rho_1)$ as a function of dimensionless time ($Jt$) for (a,d) $U/J=0.1$, (b,e) $U/J=1$ and (c,f) $U/J=10$ with two different initial conditions. For (a,b,c), both the particle is initially in a single well whereas for (d,e,f), initially one particle in each well.}
 \label{Figure 9.}
\end{figure}

In Fig.\ref{Figure 8.} we plot $S_N^{{\rm spatial}}$ as a function of dimensionless time $J t$ for both bosonic and fermions systems for no interaction  ($U=0$ and as well as with interactions ($U=J$ and $U=10 J$) for the two possible initial conditions. We show that the fermionic ``spatial'' mode-entanglement is always greater than bosonic entanglement. For both atoms initially in single well, there is only one frequency of temporal oscillation for non-interacting system. But in presence of interaction, more than one frequency show up in the dynamics. With increasing values of interactions, the bosonic and fermionic spatial-mode von Neumann entropy tend to match each other signifying similar mode entanglement property for strongly interacting system with the given initial condition. With initially two atoms in two different wells, the fermionic $S^{\rm spatial}_N$ is always greater than 1 and oscillates between 1 and 2, whereas for bosonic case it oscillates between 0 and 1.6. We find  that $S_N^{(1)}$ in the case of two component fermions varies between 1 and 2 while that in case of spinless bosons varies between 0 and 1. We find that if the system is initially prepared with both particles in same site, the temporal variation of $S_N^{(1)}$ is qualitatively similar for both spinless bosons and two-component fermions except for a shift of unity in the case of two-component fermions.  The entanglement measure $E(\rho_1)$ for N qubits is obtained by subtracting ${\rm log}_2N$ (corresponding to the local exchange correlation) from the von Neumann entropy of $\rho_1$. Therefore, for the fermions $E(\rho_1) = S_N^{(1)} - 1$ while for a pair of spinless bosons, $E(\rho_1) = S_N^{(1)}$ since for a pair of spinless bosons located at the same site there exists no exchange correlation. In Fig.\ref{Figure 9.} we have shown the time-dependence of the $E(\rho_1)$ for fermions for three different values of $U$. For spinless bosons, $E(\rho_1)$ is identical to that of fermions, however 
interpretation of entanglement in two cases is obviously different: while in case of spinless bosons, entanglement may arise only between the two spatial modes, in the case of two-component fermions or bosons the entanglement may occur in both spatial modes and spin DOF. For the other initial condition (two particles in two different sites), the value of $E(\rho_1)$ is always greater than zero. For small values of $U$, $S_N^{(1)}$ is almost always 1 for spin-less bosons and 2 for fermions.

\subsection{Fluctuation in number and phase}\label{5.2}

\begin{figure}[h]
    \centering
    \begin{minipage}{0.48\textwidth}
        \centering
        \includegraphics[width=3.2in, height=2.0in]{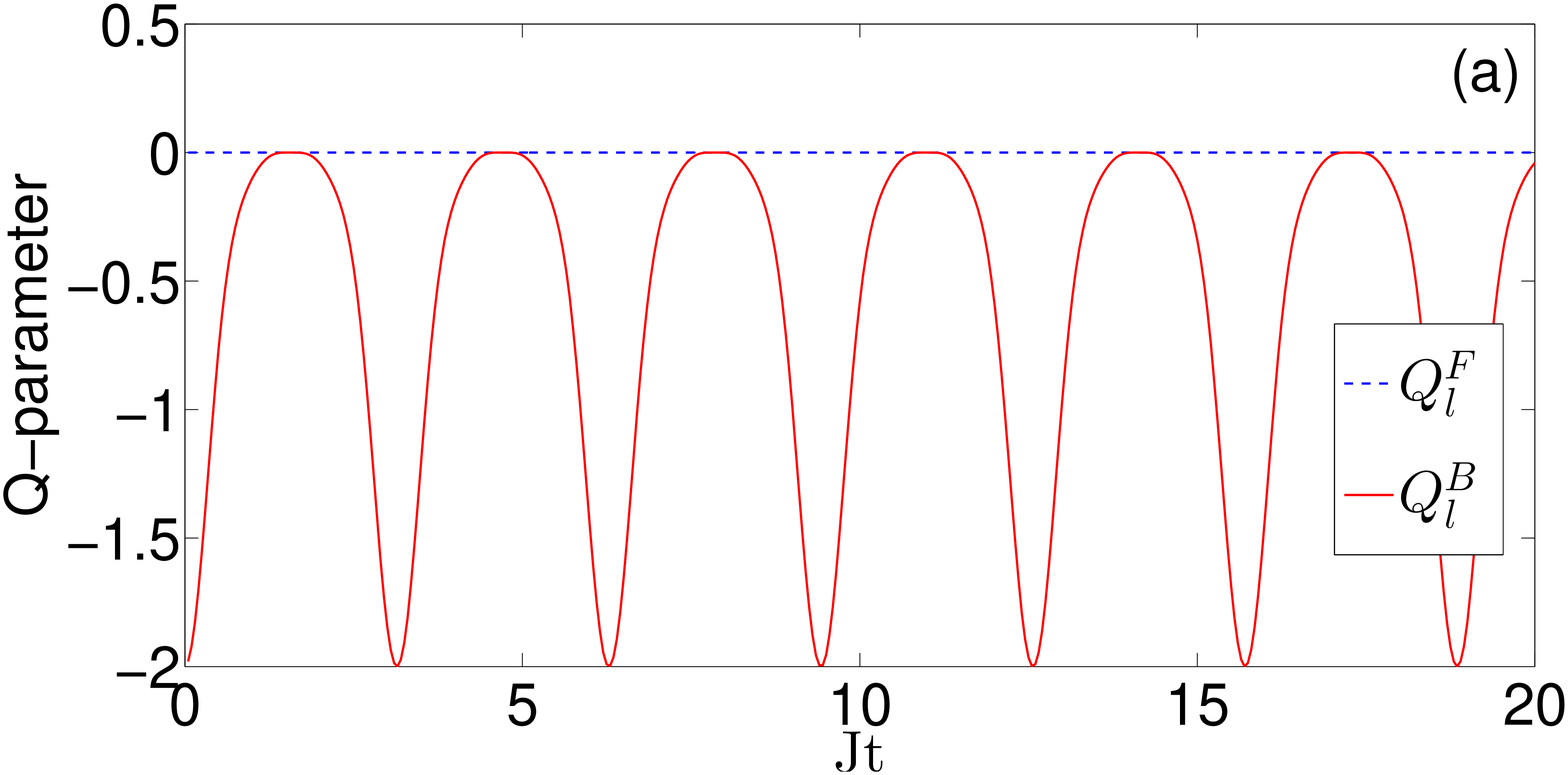}
    \end{minipage}
        \begin{minipage}{0.48\textwidth}
        \centering
        \includegraphics[width=3.2in, height=2.0in]{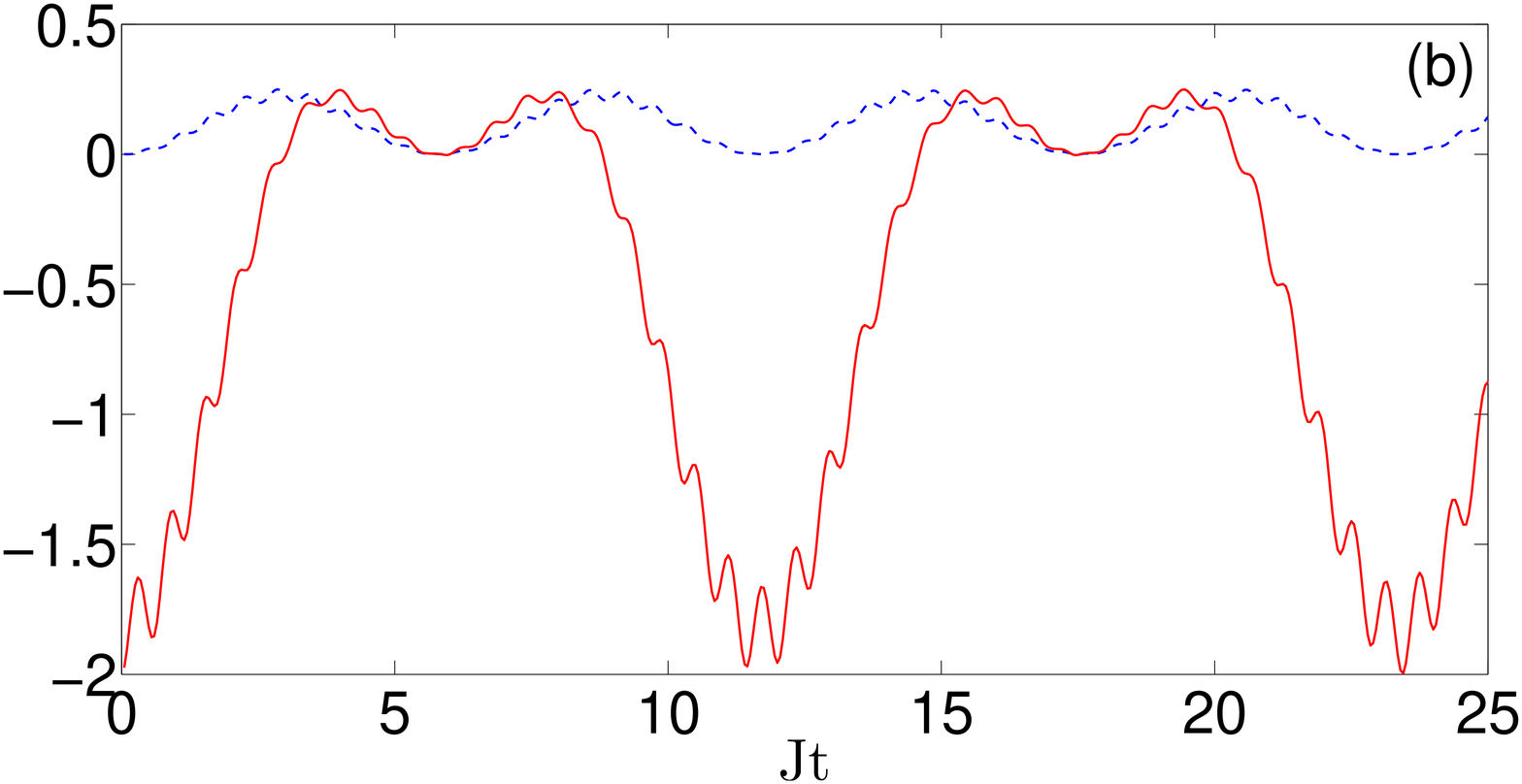}
    \end{minipage}
    \begin{minipage}{.48\textwidth}
        \centering
        \includegraphics[width=3.2in, height=2.0in]{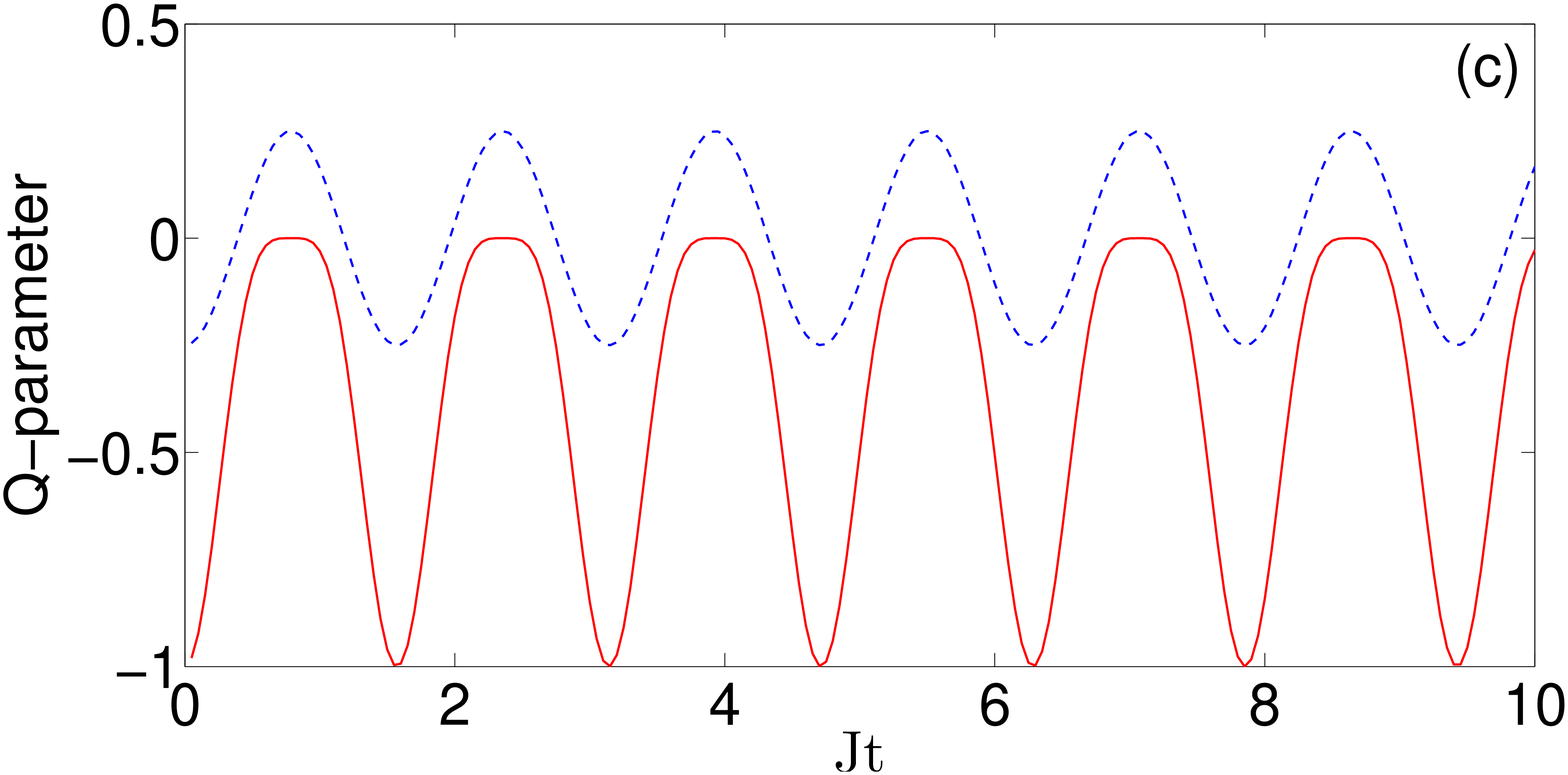}
    \end{minipage}
        \begin{minipage}{0.48\textwidth}
        \centering
        \includegraphics[width=3.2in, height=2.0in]{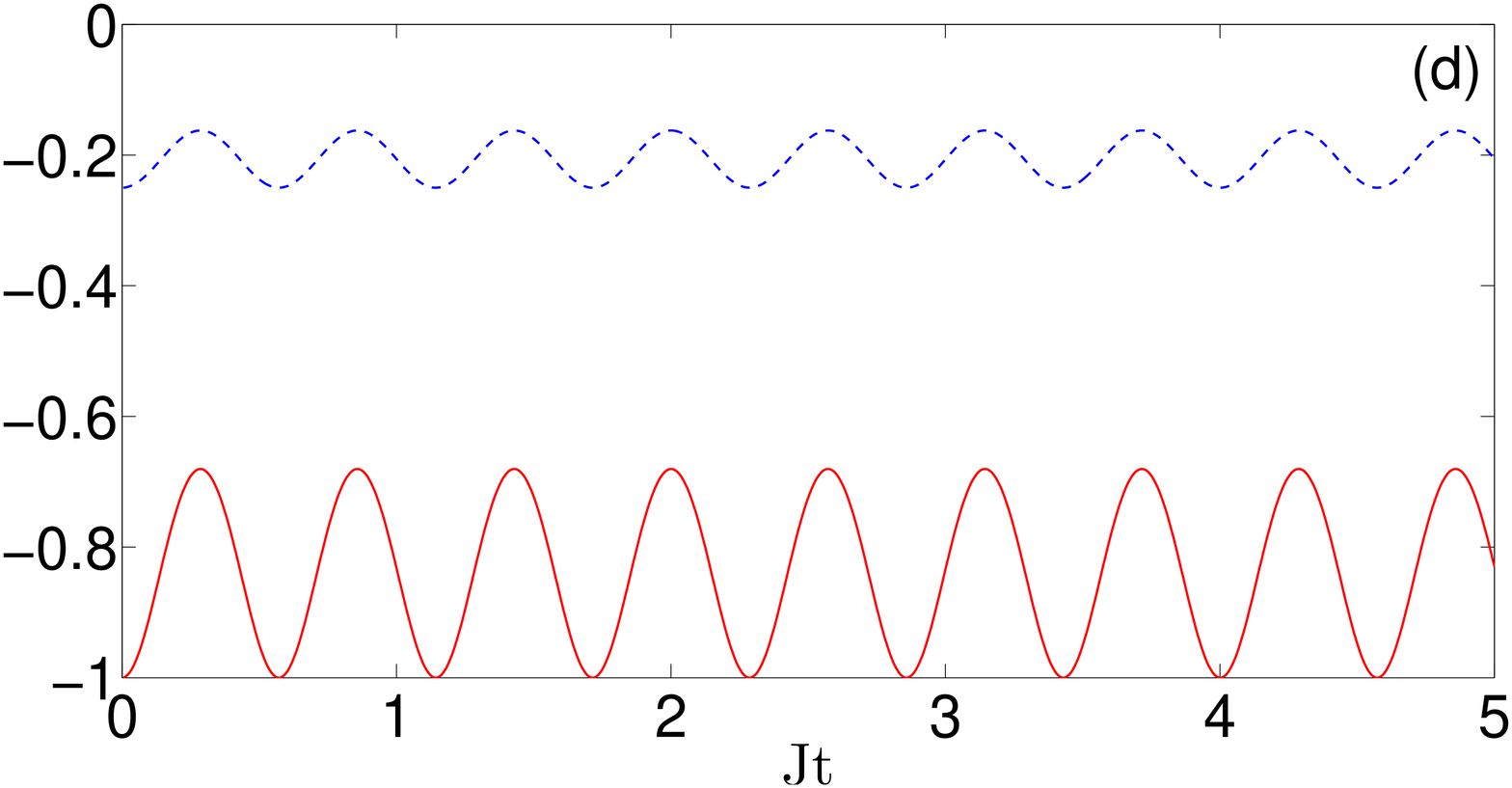}
    \end{minipage}
 \caption{\small Time evolution of fermionic (dashed) and bosonic (solid) $Q$- parameters with initially both atoms in same site (a, b) and with two atoms initially in each sites (c, d). The interaction strenth are $U=0$ (a, c) and $U\approx10J$ (b, d) with $J=150$ Hz and $a_s=52.9$nm.}
 \label{Figure 10.}
\end{figure}

We quantify the on-site number fluctuation by a parameter which is analogous to the Mandel $Q$-parameter \cite{mandel} well-known in quantum optics. 
 For the bosonic case, the on-site $Q$-parameter is defined as 
 $Q^{(B)}_j = \langle\hat{a}^\dagger_j\hat{a}^\dagger_j\hat{a}_j\hat{a}_j\rangle - \langle\hat{a}^\dagger_j\hat{a}_j\rangle^2 $ where the subscript $j$ stands for 
either $l$ (left) or $r$ (right). In terms of bosonic number operator   $\hat{N}_{j} = \hat a^{\dagger}_j \hat a_{j}$, this can be expressed as 
\begin{eqnarray}
Q^{(B)}_j =  \langle\hat{N_j}^2\rangle - \langle\hat{N_j}\rangle^2 - \langle\hat{N_j}\rangle, \hspace{1in}  
 \label{eq19}
\end{eqnarray}
From  time-dependent wave function given in Sec.\ref{3.2}, we have  $Q^{(B)}_l = 2|C_0|^2 - 4|C_0|^4  - |C_1|^4 - 4|C_0|^2|C_1|^2$. Using the Eqs.(\ref{eq12}) or Eqs.(\ref{eq13}) which correspond to initializing both atoms in the same well or  each atom in each well, respectively; an explicit expression for $Q^{{B}}_j$ can be obtained.  For the fermionic system, it is defined as $Q^{(F)}_j = \langle\hat{a}^\dagger_{j\uparrow} \hat{a}^\dagger_{j\downarrow}\hat{a}_{j\downarrow}\hat{a}_{j\uparrow}\rangle - \langle \hat{a}^\dagger_{j\uparrow} \hat{a}_{j\uparrow}\rangle \langle\hat{a}^\dagger_{j\downarrow}\hat{a}_{j\downarrow}\rangle $ which can be written as  
\begin{eqnarray}
 Q^{(F)}_j = \langle\hat{N}_{j\downarrow}\hat{N}_{j\uparrow}\rangle - \langle\hat{N}_{j\downarrow}\rangle \langle\hat{N}_{j\uparrow}\rangle 
 \label{eq18}
\end{eqnarray}
where $\hat{N}_{j\sigma} = \hat a^{\dagger}_{j\sigma} \hat a_{j\sigma}$ is the fermionic number operator. This can be calculated as  $Q^{(F)}_l = |c_0|^2|c_3|^2 - |c_1|^2|c_2|^2$.  The bosonic $Q$-parameter $Q^{(B)}_j$ has the same form as the Madel $Q$-parameter, however $Q^{(F)}_j$ has  different form and is basically the on-site two-component cross number fluctuation. However, $Q^{(B)}$ is the difference between the on-site number fluctuation and the average number. If $Q < 0$ then the fluctuation is said to be below the coherent or quantum shot noise level implying anti-bunching or non-classical behavior for the atom statistics.  For a symmetric DW, $Q^{(F)}_l=Q^{(F)}_r$. But in general, $Q^{(B)}_l\neq Q^{(B)}_r$. From symmetry, $Q^{(B)}_l = Q^{(B)}_r$ only for the condition where two bosons are initially in opposite sites but not for the other initial condition where they are initially in the same site. In Fig.\ref{Figure 10.} we have plotted $Q$-parameter of left-site mode as a function of dimensionless time $Jt$ for bosonic and fermionic system for different initial conditions. If the both atoms are initially prepared in the left well, then the temporal behavior of $Q^{(B)}_l$ and $Q^{(B)}_r$ show a shift of $\pi/2$ between them. Fig.\ref{Figure 10.} shows that when both the atoms are initially prepared in the same well (say, jth well), $Q^{(B)}_j$ for $U=0$ (non-interacting bosons) oscillates periodically between 0 and -2, while for $U\neq0$ it again oscillates periodically, albeit with modulations due to interaction effects, mostly in the negative side. This implies that for the said initial condition the bosons are mostly anti-bunched. In contrast, $Q^{(F)}_j$ for the same initial condition, is always positive for $U\neq0$ implying bunching of the two fermions, while it is always zero for $U=0$. For the other initial condition i.e., initially two atoms are in different wells, for $U=0$ the $Q$-parameter for bosons oscillates periodically 0 and -1 while that for fermions oscillates around zero equally between positive and negative sides. However, for $U/J>\!>1$  both $Q^{(B)}$ and $Q^{(F)}$ oscillate periodically entirely in the negative side implying that in both the cases the particles are anti-bunched. In the latter initial condition, temporal modulations are absent or negligible because as we learned from our studies on pair probabilities in the preceding section, the double-occupation probability ($\rho_d$) for latter initial condition with $U/J>\!>1$ is exceedingly small meaning there is hardly any effect of on-site interaction. Here inter-site and other interaction parameters are assumed to be quite small compared to $U$, otherwise those terms may cause additional modulations.

We next turn our attention to the quantum fluctuations in the phase-difference and number-difference or population imbalance between the two spatial modes. Towards this end, 
we make use of matter-wave unitary phase-difference  operators defined in Ref.\cite{jpb:2013:biswajit}. Specifically, one can define two mutually commuting phase-difference operators $\hat{C}_{lr}^{(a)}$ and $\hat{S}_{lr}^{(a)}$ corresponding to the cosine and sine, respectively,  of the phase-difference between the left and right modes of the fermions ($a \equiv F$) or bosons ($a = B$). These two operators do not commute with the population imbalance operator $\hat{W} = \hat{N}_l - \hat{N}_r$ where $\hat{N}_j = \hat{a}_j^{\dagger} \hat{a}_j $ for spinless bosons or $\hat{N}_j = \sum_{\sigma=\uparrow,\downarrow} \hat{a}_{j\sigma}^{\dagger} \hat{a}_{j\sigma} $ for two-component fermions or bosons, leading to number-phase uncertainty relations \cite{jpb:2018:kingshuk}. 
It is theoretically shown that, these  quantum phase operators are particularly important for matter-waves with low number of bosons or fermions, consistent with the similar result in case of photons as shown in \cite{Mandel:operational}. One can define an average phase fluctuation $\Delta E_{\phi} = \sqrt{(\Delta C)^2 + (\Delta S)^2}$ where $\Delta C$ and $\Delta S$ are the fluctuations corresponding to the cosine and sine of the phase-difference operators. Accordingly, standard quantum limits for number-phase uncertainty in bosonic and fermionic matter-waves are defined.   The properties of the phase fluctuations  are described in detail elsewhere \cite{jpb:2018:kingshuk}. Here we recall the salient features of bosonic and fermionic phase fluctuations: a pair of spinless bosons never exhibit any phase squeezing, in contrast a pair of 
two-component fermions shows strong phase squeezing when the two fermions are initially located in two different sites.

The phase fluctuation in the ground state $\mid a \rangle$ is analytically calculated: $\Delta E_{\phi}^F = 2\sqrt{2}J\sqrt{8J^2+(U_{-}+\Omega)^2}/[16J^2+(U_{-}+\Omega)^2]$ and $\Delta E_{\phi}^B = \sqrt{1-[32J^2(U_{-}+\Omega+ \sqrt{2}J)^2]/[16J^2 + (U_{-}+\Omega)^2]^2}$.
Average number-difference for both bosonic and fermionic cases is found to be always zero in the ground state but number fluctuations are nonzero, $\Delta N^F = 8J/\sqrt{16J^2+(U_{-}+\Omega)^2} =\Delta N^B$, where $\Delta N^{F(B)}$ stands for fluctuation in two-mode number-difference in fermionic (bosonic) case. The standard quantum limit of number-phase uncertainty for fermionic and bosonic system in the ground eigen state $|a\rangle$ is calculated to be $\Delta^F_{SQL}= \mid 16J^2_-/[16J^2_-+(U_-+\Omega)^2] \mid$ and $\Delta^B_{SQL}= \mid 4\sqrt{2}J_-[U_-+\Omega-2\sqrt{2}J_-]/[16J^2_-+(U_-+\Omega)^2] \mid$, respectively. From these expression one can infer that for $U_{-} \rightarrow \infty$, fluctuations in the ground state are $\Delta N^{F} \rightarrow 0$, $\Delta N^{B} \rightarrow 0$, $\Delta E_{\phi}^{B} \rightarrow 1/2$,  $\Delta E_{\phi}^{F} \rightarrow 0$ and $\Delta^F_{SQL}\rightarrow0$, $\Delta^B_{SQL}\rightarrow0$. However, for $U_{-} \rightarrow - \infty$, $\Delta N^{F} \rightarrow 1/2$, $\Delta N^{B} \rightarrow 1/2$, $\Delta E_{\phi}^{B} \rightarrow 3/4$,  $\Delta E_{\phi}^{F} \rightarrow 1/2$ and $\Delta^F_{SQL}\rightarrow1$, $\Delta^B_{SQL}\rightarrow1$. On the other hand, in the non-interacting limit $U_{-}\rightarrow 0$, we have $\Delta N^B \rightarrow  1/(2\sqrt{2}) $, $\Delta N^F \rightarrow  1/(2\sqrt{2})$, $\Delta E_{\phi}^F \rightarrow \sqrt{3}/4$, $\Delta E_{\phi}^B \rightarrow (1-(4+\sqrt{2})^2/32)^{\frac{1}{2}}$ and $\Delta^F_{SQL}\rightarrow1/2$, $\Delta^B_{SQL}\rightarrow1/\sqrt{2}-1/2$.

\section{Conclusions and outlook}\label{6} 

In conclusion, we have shown that two-site Hubbard models with a pair of bosonic as well as fermionic  atoms yield qualitatively same results for almost all the quantum statistical average quantities such as occupation statistics, single-particle and pair tunneling probabilities for the same input parameters. However, on-site number and inter-site quantum phase fluctuations are quite different for the two cases. Particularly interesting  is the characteristically different behavior of entanglement and phase fluctuations for a pair of two-component fermions vis-a-vis for a pair of spinless bosons. While fermions show strong phase squeezing and maximal entanglement between two spatial modes when two fermions are initially prepared in two different wells, spinless bosons never exhibit any phase squeezing and the entanglement is always less than that of fermions. So, it would be an interesting question whether there is any connection between entanglement and quantum phase fluctuations.

 Our calculations show that for appreciable finite range of interactions, one can not neglect the effects of inter-site interaction as it can significantly influence the results \cite{Yukalov:2008}. For large effective range or for a long-range interaction, the inter-site interaction $U_i$ is already found to be important \cite{Yukalov:2008}. Finite-range interactions become particularly important for magnetically tunable Feshbach resonances of ultracold atoms. Since atoms become strongly interacting near a Feshbach resonance, two-mode approximation of Hubbard model may break down \cite{Schneider,EPJ:2016}, because the on-site interaction energy $U$ then may exceed the gap between the ground and the excited bands. In that situation, one has to work out multi-band Hubbard physics as done in a recent work to show two-particle quantum correlations in higher bands of a two-site Hubbard model \cite{landman:pra:2018}. However, it follows from the theory of Jost and Kohn that, near a narrow Feshbach resonance, exploring the two-mode Hubbard physics can not be ruled out \cite{Abhik_paper}. One may also explore atom-molecule coupled  Hubbard physics with magnetically tunable Feshbach resonances for which molecular regime becomes important. Furthermore, one can  manipulate two-particle quantum correlations of a two-site Hubbard model using tunable 
 interactions \cite{landman:pra:2018}. This opens up a considerable prospect for a two-site Hubbard model as a tool for making quantum  gates \cite{Foot} and thus for exploring quantum information processing. The interatomic interaction effects on the Hubbard physics including quantum correlations can be studied either by altering scattering length or range of interaction or both. So, by building up Hubbard parameters from the solutions of a pair of atoms interacting via Jost-Kohn potentials in a DW trap, it will be possible to explore the effects  of both $a_s$ and the effective range $r_0$ on Hubbard physics. 
   
 \vspace{0.5cm}
\begin{center}
{\bf Acknowledgments}
\end{center}
One of us (SM) is thankful to the Council of Scientific and Industrial Research (CSIR), Govt. of India, for support. KA and BD thankfully acknowledge the support from the Department of Science \& Technology, Govt. of India, under the project No. SB/S2/LOP-008/2014.

\appendix
\section{Jost-Kohn potentials}\label{appendix-A}

There exists a multitude of model interaction potentials \cite{braaten:annphys:2008}, including ones that take into account effective range effects. However, Jost-Kohn potentials are least known. The three-parameter model interaction potential of Jost and Kohn \cite{JostKhon} for positive $a_s$ depends on the $s$-wave binding energy $E_b = -\hbar^2 \kappa^2/2\mu$ ($\kappa > 0$), where $\mu$ is reduced mass and
\begin{eqnarray}
 \kappa = \frac{1}{r_0}\left[1+\alpha\right]\frac{1+\Lambda}{1-\Lambda}
 \label{eq3}
\end{eqnarray}
with $-1<\Lambda<1$, $\alpha = \sqrt{1-2r_0/a_s}$ and $a_s>2r_0$. In terms of $a_s, r_0$ and $\Lambda$, the potential is 
\begin{eqnarray}
V_{+}(x) &=& V_0 e^{-2(1-\alpha) x}\frac{\Big[\big\{(1+\alpha\Lambda)(\alpha+\Lambda)(1-\alpha)(1-\Lambda^2e^{-2\beta x})\big\}^2- \Lambda^2\beta^2\big\{(1+\Lambda\alpha)^2e^{-2\alpha x} - (\alpha+\Lambda)^2e^{-2x} \big\}^2 \Big]}{\Big[(1+\alpha\Lambda)^2(\alpha+\Lambda^2e^{-2\beta x}) -(\alpha+\Lambda)^2(e^{-2(1-\alpha) x} + \alpha\Lambda^2 e^{-4x})\Big]^2}
\label{eq4}
\end{eqnarray}
where $\beta = 1+\alpha$, $V_0=\frac{4\hbar^2\alpha}{\mu r_0^2}$ and $x=\frac{r}{r_0}$ with $r$ being the inter-particle separation. The expression of model potential for negative scattering length is given by Eq.(2.29) of Ref \cite{JostKhon1}
\begin{eqnarray} 
 V_{-}(r) = - \frac{4\hbar^2}{\mu r_0^2}\frac{\alpha \beta^2\exp(-2\beta x)}{[\alpha +\exp(-2\beta x)]^2}
\label{eq5}
\end{eqnarray}

\section{Calculation of $U$ under Hubbard approximation } \label{appendix-B}
To calculate $U$,  we take the single particle 3D wave function 
\begin{eqnarray}
 \Psi(r,t) = \frac{1}{\sqrt{\pi a_\rho^2}}e^{-\frac{\rho^2}{2a_\rho^2}}\psi_{1D}(z,t) \nonumber
\end{eqnarray}
The interaction strength is calculated as
\begin{eqnarray}
 U&=&\int |\Psi(r_1)|^2V_{int}(r_1,r_2)|\Psi(r_2)|^2 d^3r_1 d^3r_2 \nonumber\\
 &=& \frac{1}{\pi^2 a_\rho^4}\int e^{-\frac{\rho_1^2+\rho_2^2}{a_\rho^2}}V_{int}(\rho_1,\rho_2,z_1,z_2)\nonumber\\&\times&\psi_{1D}(z_1)\psi_{1D}(z_2)(2\pi)^2\rho_1\rho_2 d\rho_1 d\rho_2dz_1 dz_2 \nonumber
\end{eqnarray}
Now we go to relative frame:
\begin{eqnarray}
 X=\frac{1}{2}(x_1+x_2) \hspace{1in}  x=(x_1-x_2) \nonumber\\
 Y=\frac{1}{2}(y_1+y_2) \hspace{1in}  y=(y_1-y_2) \nonumber
\end{eqnarray}
Therefore,
\begin{eqnarray}
 \rho_1^2+\rho_2^2 &=& 2(X^2+Y^2)+(x^2+y^2) \nonumber\\
 &=& 2\rho_{CM} + \frac{\rho}{2} \nonumber
\end{eqnarray}
\begin{eqnarray}
 \therefore U &=&\frac{4}{a_\rho^4}\int_0^\infty e^{-\frac{2\rho_{CM}}{a_\rho^2}}\rho_{CM} d\rho_{CM}\int \rho e^{-\frac{\rho^2}{2a_\rho^2}}V_{int}(\rho,z_1,z_2)\nonumber\\&\times&|\psi_{1D}(z_1)|^2|\psi_{1D}(z_2)|^2 d\rho dz_1 dz_2 \nonumber\\
 &=&\frac{1}{a_\rho^2}\int \rho e^{-\frac{\rho^2}{2a_\rho^2}} V_{int}(\rho,z_1,z_2) |\psi_{1D}(z_1)|^2 |\psi_{1D}(z_2)|^2 d\rho dz_1 dz_2 \nonumber
\end{eqnarray}

\end {document}